\documentclass[10pt, a4paper]{article}
\usepackage{amsmath}
\usepackage[]{graphicx}
\usepackage[pdfauthor={P. P. Cook and P. C. West},
pdftitle={G+++ and Branes}]{hyperref}
\begin{document}
\title{$\cal G^{+++}$ and Brane Solutions}
\author{Paul P. Cook\footnote{email: paul.p.cook@kcl.ac.uk} and Peter C. West\footnote{email: pwest@mth.kcl.ac.uk}\\ \and \\{\itshape Department of Mathematics, King's College London,\/}\\{\itshape Strand, London WC2R 2LS, U.K.\/}}
\begin{titlepage}
\begin{flushright}
Preprint KCL-MTH-04-06\\
{\tt hep-th/0405149}
\end{flushright}
\vspace{70pt}
\centering{\LARGE$\cal G^{+++}$ and Brane Solutions}\\
\vspace{30pt}
\def\thefootnote{\fnsymbol{footnote}}
Paul P. Cook\footnote{\href{mailto:paul.p.cook@kcl.ac.uk}{email: paul.p.cook@kcl.ac.uk}} and Peter C. West\footnote{\href{mailto:pwest@mth.kcl.ac.uk}{email: pwest@mth.kcl.ac.uk}}\\
\setcounter{footnote}{0}
\vspace{10pt}
{\itshape Department of Mathematics, King's College London,\/}\\
{\itshape Strand, London WC2R 2LS, U.K.\/}\\
\vspace{10pt}
\vspace{30pt}
\begin{abstract}
We demonstrate that the very extended $\cal G^{+++}$ group element of the form $g_A=\exp(-{\frac{1}{(\beta,\beta)}\ln N }\beta \cdot H)\exp((1-N)E_\beta)$ describes the usual BPS, electric, single brane solutions found in $\cal G^{+++}$ theories.
\end{abstract}
\end{titlepage}
\clearpage
\newpage
\section{Introduction}
Very little is known about M theory itself, but investigations of the associated supergravity theories are beginning to reveal some of its underlying symmetry. Study of the properties of the eleven dimensional supergravity theory has lead to the recent conjecture \cite{West1} that M theory possesses an $E_{11}$ Kac-Moody symmetry. Indeed it was argued \cite{West1} that an extension of eleven dimensional supergravity should possess an $E_{11}$ symmetry that was non-linearly realised. A benefit of this approach is that the symmetries found when the eleven dimensional supergravity theory was dimensionally reduced are natural occurrences of there being an $E_{11}$ symmetry in the eleven dimensional theory.\\

The same analysis was applied to the IIA and IIB supergravity theories which were conjectured to be part of a larger theory which also possessed an $E_{11}$ symmetry \cite{West1, SchnakenburgWest}. This is consistent with the idea that the type II string theories in ten dimensions and an eleven dimensional theory are part of a single theory called M theory. Indeed their common $E_{11}$ origin provides an explicit relation between the type IIB theory in ten dimensions and the eleven dimensional theory which are not related by dimensional reduction \cite{West}.\\

Arguments similar to those advocated for eleven dimensional supergravity in \cite{West1} were proposed to apply to gravity \cite{LambertWest} in D dimensions, the effective action of the closed bosonic string \cite{West1} generalised to D dimensions and the type I supergravity theory \cite{SchnakenburgWest1} and the underlying Kac-Moody algebras were identified. It was realised that the algebras that arose in all these theories were of a special kind and were called very extended Kac-Moody algebras \cite{GaberdielOliveWest}. Indeed, for any finite dimensional semi-simple Lie algebra $\cal G$ one can systematically extend its Dynkin diagram by adding three more nodes to obtain an indefinite Kac-Moody algebra denoted $\cal G^{+++}$. In this notation $E_{11}$ is written as $E_8^{+++}$. The algebras for gravity, the closed bosonic string, and type I supergravity being $A^{+++}_{D-3}$ \cite{LambertWest}, $D^{+++}_{D-2}$ \cite{West1} and $D^{+++}_8$ \cite{SchnakenburgWest1} respectively. The Dynkin diagrams for the very extended algebras considered in this paper are given in Appendix B.\\

It was proposed in \cite{West1, LambertWest, SchnakenburgWest1, GaberdielOliveWest} and \cite{EnglertHouartTaorminaWest, EnglertHouartWest, KleinschmidtSchnakenburgWest, EnglertHouart}, that the non-linear realisation of any very extended algebra ${\cal G}^{+++}$ leads to a theory, called ${\cal V}_{\cal G}$ in \cite{KleinschmidtSchnakenburgWest}, that at low levels includes gravity and the other fields and it was hoped that this non-linear realisation contains an infinite number of propagating fields that ensures its consistency. Indeed, it has been shown \cite{KleinschmidtSchnakenburgWest} that the non-linear realisation of any of the very extended algebras $\cal G^{+++}$ leads to a theory that contains gravity, gauge fields of various rank and possibly a dilaton at low levels.\\

If one dimensionally reduces a supergravity theory on a torus to $D=3$ one obtains a generic theory containing gravity, gauge fields of various rank and possibly a dilaton that is described by the properties of its scalars. A non-trivial field strength in three dimensions may have at most two indices making it the dual of the field strength derived from the scalar. In some cases these scalars belong to a non-linear realisation of some finite dimensional Lie group, $\cal G$. A set of theories whose dimensional reductions to three dimensions yielded every possible $\cal G$ was found in \cite{CremmerJuliaLuPope}. The maximally oxidised theories corresponding to $\cal G$ are those in the highest dimension, $D$, such that upon dimensional reduction to three dimensions one finds the scalars belonging to the non-linear realisation $\cal G$ \cite{CremmerJuliaLuPope}; for a detailed discussion of oxidation in the context of group theory the reader is referred to \cite{Keurentjes}. Reference \cite{LambertWest} considered the dimensional reduction of certain theories containing gravity, gauge fields of various rank and a dilaton and found that only for very specific couplings did the scalars belong to a non-linear realisation of $\cal G$ and in all cases these couplings were oxidised theories. Indeed a correspondence between the low-level fields in the adjoint representation of $\cal G^{+++}$ and the maximally oxidised theory associated with $\cal G$ has been demonstrated in reference \cite{KleinschmidtSchnakenburgWest}.\\

There is a very substantial literature on brane solutions in supergravity and, indeed, in a generic theory consisting of gravity, gauge fields of various rank and a dilaton we will find that our equation (\ref{groupelement}) encodes the electric brane solutions found in these theories \cite{DuffLu, EnglertHouartArgurio}. The solutions for the oxidised theories mentioned above were considered in \cite{EnglertHouartWest}.\\

It was shown in \cite{West} that the usual BPS solutions of the type IIA and type IIB ten dimensional supergravity theory and the eleven dimensional theory possess a general formulation in terms of $E_{11}$ group elements, namely, $\exp(-\frac{1}{2}\ln N \beta \cdot H)\exp((1-N)E_\beta)$. The form of this expression makes it a simple process to write down any solution from a particular $E_{11}$ group element, and vice-versa.\\

In this paper we generalise the result of \cite{West} so that it may be applied to any $\cal G^{+++}$ algebra. We find that the group element takes the general form
\begin{equation}
g_A=\exp(-{\frac{1}{(\beta,\beta)}\ln N }\beta \cdot H)\exp((1-N)E_\beta)
\label{groupelement}
\end{equation}
Where $(\beta,\beta)$ is the norm squared of the root whose corresponding generator is $E_\beta$, when the squared length of the roots on the gravity line have been normalised to two. For the cases where $(\beta,\beta)=2$ the group element becomes that given in \cite{West}. $N$ is a harmonic function taking the form
\begin{eqnarray}
N=1+\frac{k}{r^{[(D-2)-2 \sum \limits_{i\in GL} i\sum \limits_{j\notin GL}{\rho_{ij}(\alpha_j,\beta)}-\sum \limits_{j\notin GL}D]}}
\label{harmonicfunction}
\end{eqnarray}
Where $GL$ refers to nodes of the Dynkin diagram found on the gravity line of each theory, $\rho_{ij} = \frac{(\alpha_i,\alpha_j)}{(\alpha_i,\alpha_i)(\alpha_j,\alpha_j)}$ is the metric of the Cartan subalgebra, $D$ is the dimension of spacetime and $r^2 = \delta^{ab}x_bx^a$ is the radial extension in the transverse coordinates, $x^a$.\\

In section 2 we demonstrate the use of equation (\ref{groupelement}) for the simply-laced group $D_{n-3}^{+++}$ and deduce the usual BPS solutions, while in section 3 we apply the method to the non-simply laced groups $B_{n-3}^{+++}$ and $F_4^{+++}$. Each of the $\cal G^{+++}$ theories we consider in this paper has a $pp$-wave solution arising from the gravity sector but their derivation does not differ significantly from those given in \cite{West} and are not calculated in this paper. Finally, in section 4 we consider some of the consequences of this result in relation to brane solutions emerging from the higher order content of the $\cal G^{+++}$ theories. In Appendix A we give the application of the solution to the remaining very extended groups, with the exception of $C_{n-3}^{+++}$\footnote{The case of $C_{n-3}^{+++}$ is very cumbersome: to consider the embedding of an $A_3$ gravity line we must delete no less than $(n-3)$ nodes, which gives rise to $(n-4)^2$ scalar fields and $2n-8$ one-forms \cite{KleinschmidtWest}, while we offer no proof of the element (\ref{groupelement}) for $C_{n-3}^{+++}$, we have no reason to believe it will fail in this theory}, and the relevant very extended Dynkin diagrams are given in Appendix B.
\subsection{Kac-Moody Algebras}
A Kac-Moody algebra is constructed from its Cartan matrix, $A_{ab}$\footnote{For a review of Cartan's classification of the finite semisimple Lie groups $\cal G$ and the corresponding Cartan matrices, $A_{ab}$, see \cite{Cahn, Georgi}.} and the Chevalley generators. A Cartan matrix $A_{ab}$ is given in terms of the simple roots, $\alpha_a$, of an algebra as
\begin{equation}
A_{ab}=2\frac{(\alpha_a,\alpha_b)}{(\alpha_a,\alpha_a)}
\end{equation}

Where $(\alpha_a,\alpha_b)$ is the Kac-Moody generalised Killing form acting on the corresponding elements of the Cartan sub-algebra. The diagonal elements of $A_{ab}$ are all 2 and the off-diagonal elements are negative integers, with the zeroes being symmetrically distributed such that
\begin{equation}
A_{ab}=0\qquad \Leftrightarrow \qquad A_{ba}=0
\end{equation}

The generators associated with positive simple roots are labeled $E_a=E_{\alpha_a}$ and those associated to negative simple roots $F_a \equiv E_{-{\alpha_a}}$. A Kac-Moody algebra is then uniquely constructed from the commutators of $E_a$, $F_a$ and the Cartan sub-algebra generators, $H_a$, subject to the following Serre relations
\begin{align}
\nonumber [H_a, H_b]&=0\\
[H_a, E_b]=A_{ab}E_b \qquad & \qquad [H_a, F_b]=-A_{ab}F_b \label{Serre}\\
\nonumber [E_a, F_b]&= \delta_{ab}H_b\\
\nonumber [E_a, \ldots [E_a,E_b]]=0 \qquad & \qquad [F_a, \ldots [F_a,F_b]]=0
\end{align}

In the final line there are $(1-A_{ab})$ $E_a$'s in the first equation and the same number of $F_a$'s in the second equation.\\

The Kac-Moody algebra may also be formulated in the Cartan-Weyl basis from $H_i$, $E_a$ and $F_a$, where $H_i$ is related to the Cartan sub-algebra by
\begin{align}
\nonumber H_a&=2\frac{\alpha_a^iH_i}{(\alpha_a,\alpha_a)}\\
\Rightarrow \alpha_a \cdot H &=\frac{(\alpha_a,\alpha_a)}{2}H_a
\label{Cartanbasis}
\end{align}
\subsection{Constructing the Algebra}
Any $\cal G^{+++}$ belongs to a class of Lorentzian Kac-Moody algebras considered in \cite{GaberdielOliveWest}. These algebras have the property that their associated Dynkin diagram possess at least one node which upon the deletion leads to component diagrams which are of finite type except for at most one of affine type. The authors of \cite{GaberdielOliveWest} have studied the properties of the Lorentzian algebras in terms of the algebra associated with the Dynkin diagram that remains after the node deletion. In this paper we may sometimes wish to delete more than one node from some Dynkin diagrams in such a way that the remaining diagram is an $A_n$ diagram. The remaining $A_n$ sub-algebra gives rise to the gravity sector and is indicated in Appendix B on the Dynkin diagrams for the relevant $\cal G^{+++}$ by solid nodes, we refer to these roots as the gravity line. We use this remaining sub-algebra to investigate the Kac-Moody algebra, $\cal G^{+++}$. Our generators therefore appear in an $SL(n)$ sub-algebra and are classified by the level at which they arise in the decomposition from the very extended algebra. In this paper the level is the coefficient, or coefficients, of the simple roots that are eliminated from the very extended algebra to obtain a finite $SL(n)$ sub-algebra. Any root, $\alpha$, may be expressed as a sum of simple roots $\alpha={\sum_a}n_a\alpha_a$ where $a$ runs over the simple roots of the very extended algebra. If the $SL(n)$ sub-algebra is formed by removing simple root $c$ then $\alpha$ is a level $n_c$ root; if roots $c$ and $d$ are removed then $\alpha$ has level $(n_c,n_d)$ and so on.\\

By definition a Kac-Moody algebra is the multiple commutators of the simple root generators, $E_a$ and separately those of $F_a$, subject to the Serre relations (\ref{Serre}). However as a matter of practise this is very difficult to carry out, in this section we explain how to construct the $\cal G^{+++}$ algebra using the tables of low-order generators in \cite{KleinschmidtSchnakenburgWest}. If we consider two $A_n$ representations with root coefficients $(a_1a_2\ldots a_n)$ and $(b_1b_2\ldots b_n)$ then their commutator has a root which is the sum, i.e. it has root coefficients $((a_1+b_1)(a_2+b_2)\ldots (a_n+b_n))$.\\

The Borel sub-algebra is formed from the Cartan sub-algebra generators, $H_a$, where $a=1\ldots n$, the positive root generators, ${K^a}_b$, where $a<b$ and $a,b=1\ldots n $, and the generators, ${R^{a_1\ldots a_n}}_{b_1\ldots b_m}$ which arise from our decomposition of $\cal G^{+++}$. The generators of the Cartan sub-algebra are constructed from the ${K^a}_a$ generators where $a=1\ldots n$ and a dilaton generator, which we label $R_0$, although some of the theories we are interested in have no so-called "dilaton". The dilaton generator is so called because it leads to a dilaton scalar field in a non-linear realisation. The ${K^a}_b$ and ${R^{a_1\ldots a_n}}_{b_1\ldots b_m}$ obey the commutation rules
\begin{align}
\nonumber[{K^a}_b,{K^c}_d]=& \delta^b_c{K^a}_d-\delta^a_d{K^b}_c\\
[{K^a}_b,{R^{c_1\ldots c_n}}_{d_1\ldots d_m}]=& \delta^{c_1}_b{R^{ac_2\ldots c_n}}_{d_1\ldots d_m}+\ldots +\delta^{c_n}_b{R^{c_1\ldots c_{n-1}a}}_{d_1\ldots d_m}\\
\nonumber &-\delta^a_{d_1}{R^{c_1\ldots c_n}}_{bd_2\ldots d_m}-\ldots -\delta^a_{d_m}{R^{c_1\ldots c_n}}_{d_1\ldots d_{m-1}b}
\end{align}

Where the second equation includes a contribution from the trace of ${K^a}_b$, which is really a $GL(n)$ generator.\\

Let us make use of the following notation for a generic commutator
\begin{equation}
[R^{n_1n_2\ldots n_a}_{(s_1)},R^{n_1n_2\ldots n_b}_{(s_2)}]=c^{s_1,s_2}_{a,b}R^{n_1n_2\ldots n_{a+b}}_{(s_1+s_2)}
\label{generalcommutator}
\end{equation}
The $s$ labels are the levels associated with the elimination of one of the simple roots in theories where the decomposition involves the elimination of more than one simple root. It is used to differentiate different types of generators, and can be simply read off from the root. We define which root is associated with the $s$ labels in each of the $\cal G^{+++}$ that we consider. In $\cal G^{+++}$ theories where only one root is eliminated the $s$ label is redundant and will be dropped.\\

Some general restrictions on $c^{s_1,s_2}_{a,b}$ are determined from the general Jacobi identity, where we simplify our notation so that the figure in square brackets is the number of antisymmetrized indices on each operator
\begin{align}
\nonumber &[R^{[a]}_{s_1},[R^{[b]}_{s_2},R^{[c]}_{s_3}]]+[R^{[b]}_{s_2},[R^{[c]}_{s_3},R^{[a]}_{s_1}]]+[R^{[c]}_{s_3},[R^{[a]}_{s_1},R^{[b]}_{s_2}]]=0\\
&\Rightarrow c^{s_1,s_2+s_3}_{a,b+c}c^{s_2,s_3}_{b,c}+c^{s_2,s_3+s_1}_{b,c+a}c^{s_3,s_1}_{c,a}+c^{s_3,s_1+s_2}_{c,a+b}c^{s_1,s_2}_{a,b}=0
\label{jacobiidentity}
\end{align}
Having commenced with the generators associated with the simple roots, we construct new generators by taking their commutators as we proceed. Subsequently by taking a commutator with a third generator we find restrictions on our commutator coefficients, $c^{s_1,s_2}_{a,b}$, using the Jacobi identity given above.\\

We will be particularly interested in finding how the dilaton commutes with other generators, which is denoted by $R^{[a]}_{s_1}=R_0$ to be the dilaton generator. The dilaton generator appears as a member of the Cartan sub-algebra, so it doesn't change the $A_n$ representation of the operator it commutes with. Of particular use are the following specific cases of (\ref{jacobiidentity}) which give the commutator coefficients for the general commutator $[R_0, R^{[m]}_s]=c_{0,m}^{0,s}R^{[m]}_s$ in a recursive form which we have found by setting $b=1$ and $c=m-1$ for the following values of $s_2$ and $s_3$ in (\ref{jacobiidentity})
\begin{align}
s_2=0, s_3=0 \qquad\Rightarrow\qquad c_{0,m}^{0,0}&=c_{0,m-1}^{0,0}+c_{0,1}^{0,0} \label{jicoefficients1}\\
s_2=0, s_3=1 \qquad\Rightarrow\qquad c_{0,m}^{0,1}&=c_{0,m-1}^{0,1}+c_{0,1}^{0,0}\\
s_2=1, s_3=0 \qquad\Rightarrow\qquad c_{0,m}^{0,1}&=c_{0,m-1}^{0,0}+c_{0,1}^{0,1}\\
s_2=1, s_3=1 \qquad\Rightarrow\qquad c_{0,m}^{0,2}&=c_{0,m-1}^{0,1}+c_{0,1}^{0,1} \label{jicoefficients4}
\end{align}
These relations allow us to find all the commutators with the dilaton, $R_0$, up to low orders, and we only need to specify one unknown which we choose to be ($c_{0,1}^{0,0}$).\footnote{The relation between $c_{0,1}^{0,1}$ and $c_{0,1}^{0,0}$ can be found by considering the commutation any of the Cartan elements, e.g. $H_n$, that contain $R_0$ with the generator for the simple root corresponding the second eliminated root, e.g. for the $F_4^{+++}$ algebra this is $E_7$, which can be seen in Appendix B on the Dynkin diagram.} Similar recursive commutator relations can be found for higher order generators by considering $s$ labels which are greater than one in equation (\ref{jacobiidentity}).\\

Let us consider the example of $F_4^{+++}$ to demonstrate how we build the algebra. The low order generators of $F_4^{+++}$ are given in reference \cite{KleinschmidtSchnakenburgWest} and we list them here with their associated root coefficients in brackets after them. 
\begin{alignat}{4}
\nonumber & &\qquad &R_0 (0000000)& \qquad &R_1 (0000001)& \qquad & \\ 
\nonumber &&\qquad &R_0^6 (0000010) & \qquad & R_1^6 (0000011) & \qquad & \\ 
& &\qquad &R_0^{56} (0000120) & \qquad &R_1^{56} (0000121)& \qquad &R_2^{56} (0000122) \\ 
\nonumber & &\qquad & & \qquad &R_1^{456} (0001231) & \qquad &R_2^{456} (0001232) \\ 
\nonumber & &\qquad & & \qquad &R_1^{3456} (0012341) & \qquad &R_2^{456,6} (0001242)\\
\nonumber & &\qquad & & \qquad & & \qquad &R_2^{3456} (0012342) 
\end{alignat}
We make use of this bracket notation throughout this paper, it is simply a list of the coefficients $n_a$ when a root, $\alpha$, associated with a particular generator, is written as a positive sum of simple roots $\alpha=\sum_a{n_a\alpha_a}$, for example $(0000121)$ corresponds to the root $\alpha=\alpha_5+2\alpha_6+\alpha_7$.\\

The seven generators associated with the simple roots are $E_a={K^a}_{a+1}$ for $a=1\ldots 5$, $E_6=R_0^6$ and $E_7=R_1$. The subscript $s$ labels are the levels corresponding to the simple root $R_1$. We read off the following commutation relations for the dilaton $R_0$ with the other generators from the root coefficients
\begin{alignat}{3}
\nonumber &[R_0,R_0^a]= c_{0,1}^{0,0}R_0^a& \qquad &[R_0,R_1^a]= c_{0,1}^{0,1}R_1^a& \qquad &\\
\nonumber &[R_0,R_0^{ab}]= c_{0,2}^{0,0}R_0^{ab}& \qquad &[R_0,R_1^{ab}]= c_{0,2}^{0,1}R_1^{ab} &\qquad &[R_0,R_2^{ab}]= c_{0,2}^{0,2}R_2^{ab} \\
& &\qquad &[R_0,R_1^{abc}]= c_{0,3}^{0,1}R_1^{abc} &\qquad &[R_0,R_2^{abc}]= c_{0,3}^{0,2}R_2^{abc} \\
\nonumber & &\qquad &[R_0,R_1^{abcd}]= c_{0,4}^{0,1}R_1^{abcd} &\qquad &[R_0,R_2^{abc,d}]= c_{0,4}^{0,2}R_2^{abc,d}
\end{alignat}
By considering $[H_7,E_7]=2E_7$ and $[H_7,E_6]=-E_6$, from equation (\ref{Serre}), we find that $c_{0,1}^{0,1}=-c_{0,1}^{0,0}$ and using equations (\ref{jicoefficients1})-(\ref{jicoefficients4}) we find that in terms of $c_{0,1}^{0,0}$ the commutation relations above are
\begin{alignat}{3}
\nonumber &[R_0,R_0^a]= c_{0,1}^{0,0}R_0^a& \qquad &[R_0,R_1^a]= -c_{0,1}^{0,0}R_1^a& \qquad &\\
\nonumber &[R_0,R_0^{ab}]= 2c_{0,1}^{0,0}R_0^{ab}& \qquad &[R_0,R_1^{ab}]= 0 &\qquad &[R_0,R_2^{ab}]= -2c_{0,1}^{0,0}R_2^{ab} \\
& &\qquad &[R_0,R_1^{abc}]= c_{0,1}^{0,0}R_1^{abc} &\qquad &[R_0,R_2^{abc}]= -c_{0,1}^{0,0}R_2^{abc} \\
\nonumber & &\qquad &[R_0,R_1^{abcd}]= 2c_{0,1}^{0,0}R_1^{abcd} &\qquad &[R_0,R_2^{abc,d}]= 0
\end{alignat}
We fix $c_{0,0}^{0,1}$ to be $\frac{1}{\sqrt{8}}$ and obtain the relations given later in equation (\ref{Fcommutators}). Other commutators can easily be found using the table of roots given in \cite{KleinschmidtSchnakenburgWest} together with the Jacobi identity (\ref{jacobiidentity}) and the Serre relations (\ref{Serre}).
\subsection{Highest and Lowest Weights in $\cal G^{+++}$ Theories}
In any finite dimensional semi-simple Lie algebra $\cal G$ the highest weight $\Lambda_H$ and lowest weight $\Lambda_L$ of any representation are related by the formula $\Lambda_L=S_0 \Lambda_H$ where $S_0$ is the longest element of the Weyl group of $\cal G$. We recall that any Weyl group element $S$ can be expressed as a product of Weyl group elements $S_a=S_{\alpha_a}$ corresponding to the reflections in the simple roots $\alpha_a$ of $\cal G$. There are a number of ways to write any $S$ involving different
numbers of $S_a$, but there is obviously always a way that involves the smallest number of $S_a$'s.  This smallest number of $S_a$'s required is called the length of the Weyl group element $S$. It turns out that the Weyl group element with the maximum length is unique and this is the one that relates the highest and lowest weights. The Weyl group element $S_b$ acts on the simple roots by 
\begin{equation}
S_b\alpha_a=(s)_{a}{}^b\alpha_b=\alpha_a-2{(\alpha_b,\alpha_a)\over(\alpha_b,\alpha_b)}\alpha_b
\end{equation}
and on the elements of the Cartan sub-algebra $H_a$ by $S_b H_a=(s)_{a}{}^bH_b$. Hence the transformation of the the Cartan sub-algbra elements by the longest element of the Weyl group is given by $S_0H_a=(s_0)_{a}{}^bH_b$\\

For $A_n$ the action of the Weyl group  on the Cartan sub-algebra is given by 
\begin{equation}
S_a H_b=H_b \ b\not=a,\ b\not=a\pm 1,\ S_nH_n=-H_n,\ S_aH_{a\pm 1}=H_a+H_{a+1}
\end{equation}
In terms of the generators $K^a{}_b$ of $A_n$ the Cartan sub-algebra elements are given by $H_a= K^a{}_{a}-K^{a+1}{}_{a+1}$ and the effect of the Weyl group element $S_a$ is 
\begin{equation}
K^a{}_{a}\ \leftrightarrow \ K^{a+1}{}_{a+1}
\end{equation}
all other elements being left inert. The longest Weyl group element of $A_n$ is given by 
$S_0=(S_1\ldots S_n)(S_1\ldots S_{n-1})\ldots (S_1S_2)(S_1)$.\\

In this paper we will consider the generators of $G^{+++}$ decomposed level by level in terms of representations of $A_n$ for suitable $n$. The highest $A_n$ weight generators in $\cal G^{+++}$ and their $\cal G^{+++}$ roots are listed in reference \cite{KleinschmidtSchnakenburgWest}. The highest $A_n$ weight generators have $SL(n+1)$ indices that that are the highest possible. For example, 
the generator $R^a$ belong to the one rank tensor representation of $SL(n+1)$ and the generator corresponding to the  highest $A_n$ weight is $R^{n+1}$. The generator corresponding to lowest
$A_n$ weight is proportional to $R^1$ and one can verify that the highest and lowest $A_n$ weights are indeed related by the action of $S_0$.\\

As we are dealing with a non-linear realisation each generator in Borel sub-algebra is associated with a field. However, we will be interested in electric branes which couple to fields with a 1 index and so correspond to  lowest $A_n$ weight generators in $G^{+++}$. In equation \ref{groupelement} we will also require the $\cal G^{+++}$ root and associated Cartan sub-algebra element that corresponds
to the lowest weight of $A_n$. We can find this using the action of $S_0$ as described above. The reader may verify that this has the same effect as applying $K^a{}_b$ generators to the generators of $G^{+++}$ to go between the highest and lowest $A_n$ weight generators. 

\subsection{Solutions in Generic Gravity Theories}
There is a very substantial literature on brane solutions in generic supergravity theories \cite{DuffLu}, for a review see \cite{Argurio, Stelle}. We use the notation of \cite{Argurio} to express the general equations of motion, line element and dilaton that arise from a theory containing gravity, a gauge field and a dilaton, which is the truncation of the full supergravity action, and allows us to investigate single brane solutions. The generic action integral is
\begin{equation}
A=\frac{1}{16\pi G_D}\int {d^Dx\sqrt{-g}(R-\frac{1}{2}\partial_\mu\phi\partial^\mu\phi-\frac{1}{2.n_i!}e^{a_i\phi}F_{a_1\ldots a_{n_i}}F^{a_1\ldots a_{n_i}})}
\end{equation}
Where $D$ is the dimension of the background spacetime, $a_i$ is the dilaton coupling constant, $F_{a_1\ldots a_{n_i}}$ is a general $n_i$-form field strength. We note here that the equations of motion derived from a truncated action, as above, is always consistent with the single electric brane solutions presented here. Chern-Simons terms in the full action will alter the equation of motion obtained from varying the gauge potentials. For our solutions the only non-zero gauge potentials possess a timelike index, any remaining Chern-Simons-like terms in the gauge equation  will be a wedge product of such gauge fields and so are identically zero. Consequently the Chern-Simons terms do not effect our single electric brane solutions as they vanish at the level of the equations of motion. The equations of motion coming from the general action above are
\begin{align}
\nonumber {R^\mu}_\nu \nonumber =& \frac{1}{2}\partial^\mu\phi\partial_\nu\phi+\frac{1}{2n_i!}e^{a_i\phi}(n_iF^{\mu a_2\ldots a_{n_i}}F_{\nu a_2\ldots a_{n_i}} -\frac{n_i-1}{D-2}{\delta^\mu}_\nu F_{a_1\ldots a_{n_i}}F^{a_1\ldots a_{n_i}})\\
&\partial_\mu(\sqrt{-g}\partial^\mu\phi)=\frac{\sqrt{-g}.a_i}{2.n_i!}e^{a_i\phi}F_{a_1\ldots a_{n_i}}F^{a_1\ldots a_{n_i}} \\
\nonumber &\partial_\mu(\sqrt{-g}e^{a_i\phi}F^{\mu a_2 \ldots a_{n_i}})= 0
\end{align}
A theory containing such field strengths, $F_{a_1\ldots a_{n_i}}$, has conserved charges which are associated with $(n_i-2)$-branes, and their dual field strengths are associated to $(D-n_i-2)$-branes. A BPS $p$-brane solution with arbitrary dilaton coupling has line element
\begin{equation}
ds^2=N_p^{-2(\frac{D-p-3}{\Delta})}(-dy_1^2+dx_2^2+\ldots dx_{p+1}^2)+N_p^{2(\frac{p+1}{\Delta})}(dx_{p+2}^2+\ldots dx_D^2) \label{Lineelement}
\end{equation}
Where $\Delta=(p+1)(D-p-3)+\frac{1}{2}a_i^2(D-2)$, $a_i$ is the dilaton coupling constant of the associated field strength, $F_{a_1\ldots a_{n_i}} 
$, and $N_p$ is an harmonic function, equal to (\ref{harmonicfunction}), and taking the general form
\begin{equation}
N_p=1+\frac{1}{D-p-3}\sqrt{\frac{\Delta}{2(D-2)}}\frac{\|\bf Q\|}{r^{(D-p-3)}} 
\end{equation}
Where $\bf Q$ is the conserved charge associated with the $p$-brane solution, and $r^2={x_{p+2}^2+\ldots x_D^2}$. We are able to read from any given line element the value of the dilaton coupling constant, $a_i$, to within a minus sign. The associated dilaton, for coupling constant $a_i$, is
\begin{equation}
e^{\phi}=N_p^{a_i\frac{D-2}{\Delta}} \label{Dilaton}
\end{equation}
A second brane solution is associated with the dual of the field strength that gives rise to the $p$-brane solution. This is a $(D-p-4)$-brane, where the dilaton coupling constant, $-a_i$, is now the negative of that for the $p$-brane. We find that for the $(D-p-4)$-brane
\begin{equation}
\Delta'=(D-p-3)(p+1)+\frac{1}{2}a_i^2(D-2)=\Delta
\end{equation}
And the line element for the brane associated with the dual field strength is
\begin{equation}
ds^2=N_{D-p-4}^{-2(\frac{p+1}{\Delta})}(-dy_1^2+dx_2^2+\ldots dx_{D-p-3}^2)+N_{D-p-4}^{2(\frac{D-p-3}{\Delta})}(dx_{D-p-2}^2+\ldots dx_D^2) \label{Duallineelement}
\end{equation}
The associated dilaton is
\begin{equation}
e^{\phi}=N_{D-p-4}^{-a_i\frac{D-2}{\Delta}} \label{Dualdilaton}
\end{equation}
Ensuring that the line elements are related as (\ref{Lineelement}) is to (\ref{Duallineelement}) and confirmation of a change of sign in the power of the harmonic function $N$ in the dilaton, as between (\ref{Dilaton}) and (\ref{Dualdilaton}), enables us to recognise which field strengths are dual to each other in the $\cal G^{+++}$ considered in this paper.

\subsection{The Dilaton Generator}
Equation (\ref{groupelement}) plays a central role in our work and may contain the dilaton generator, $R_0$, through $\beta \cdot H$ for particular algebras. The non-linear realisation allows us to write down the most generally covariant field equations for the field content arising from a particular $\cal G^{+++}$ in which the dilaton field, $A$ appears within a factor in the field strength. The procedure for finding the field equations is given in \cite{SchnakenburgWest, West1, West2} and, using the above commutator relations, in the non-linear realisation of $\cal G^{+++}$ the field strengths take the form

\begin{equation}
{F_{a_1a_2\ldots a_p}}_{s_1}=pe^{-c_{0,p-1}^{0,s_1}A}(\partial_{[a_1}{A_{a_2\ldots a_p]}}_{s_1}+\ldots)
\label{generalcovariantfieldstrength}
\end{equation}

Where "$+\ldots$" indicates terms which are not total derivatives but have the correct number of indices and within which the sum of $s$ labels matches the $s$ label of the field strength.\\

The required dual field strength in the non-linear realisation is found to take the form
\begin{equation}
{F_{a_1a_2\ldots a_{(D-p)}}}_{s_2}=(D-p)e^{-c_{0,D-(p-1)}^{0,s_2}A}(\partial_{[a_1}{A_{a_2\ldots a_{D-p}]}}_{s_2}+\ldots)
\end{equation}

Where D is the number of background spacetime dimensions, and we construct the remaining field strengths similarly. Having constructed the field strengths for a particular theory, we follow the process in \cite{SchnakenburgWest, West1, West2} and write down the equations relating a field strength and its Hodge dual, which take the general form

\begin{equation}
*{F_{a_1a_2\ldots a_{(D-p)}}}_{s_1}\equiv{F_{a_1a_2\ldots a_p}}_{s_1}=\frac{1}{p!}\epsilon_{a_1a_2\ldots a_D}F_{s_2}^{a_{(p+1)}\ldots a_D}
\label{Hodgedual}
\end{equation}

Where the Hodge dual is indicated by *, and $\epsilon_{a_1a_2\ldots a_D}$ is the usual completely antisymmetric tensor.\\

We then substitute our expressions for ${F_{a_1a_2\ldots a_p}}_{s_1}$ and ${F_{a_1a_2\ldots a_{D-p}}}_{s_2}$ into this equation. The second order field equations are obtained by bringing all factors containing the dilaton field $e^{kA}$ together as a coefficient of the field strength appearing in the Lagrangian and differentiating the equation. Any total derivatives vanish by the Bianchi identity and we obtain an equation of the form

\begin{equation}
\partial_{[a_{m}}*(pe^{(-c_{0,p-1}^{0,s_1}+c_{0,D-(p-1)}^{0,s_2})A}({\partial_{a_1}A_{a_2\ldots a_p]}}_{s_1})+\ldots)=\partial_{[a_{m}}(+\ldots)_]
\end{equation}

The "$+\ldots$" correspond to the non-total derivative terms coming from our non-linear formulation of the field strengths. The coefficient of the total derivative on the left-hand-side is now exactly that which appears in a Lagrangian formulation of this system, where the field strength is just the total derivative, $p\partial_{[a_1}A_{a_2\ldots a_p]}$. These considerations are useful in making comparisons with the literature, in particular with \cite{CremmerJuliaLuPope}. The reader may see a fully worked example of this method for $D_{n-3}^{+++}$ in section 2.\\

\section{Simply Laced Groups}
A simply laced group has simple roots which all have the same length, here our roots are normalised such that $(\alpha_a, \alpha_a)=2$.
\subsection{Very Extended $D_{n-3}$}
$D_{24}^{+++}$ is the symmetry underlying the effective action of the bosonic string theory \cite{West1}, and the analogous n-dimensional generalisation is $D_{n-3}^{+++}$ \cite{West1, EnglertHouartTaorminaWest}, these are maximally oxidised theories. The Dynkin diagram for $D_{n-3}^{+++}$ is shown in Appendix B, where the solid nodes indicate the gravity line we shall consider, which is an $A_{n-2}$ sub-algebra.\\

We decompose the $D_{n-3}^{+++}$ algebra with respect to its $A_{n-2}$ sub-algebra, the simple roots whose nodes we delete are $\alpha_{n-1}$ and $\alpha_n$, as enumerated in Appendix B and we associate the levels $l_1$ and $l_2$ with these respectively. Our $s$ labels are chosen to be the $l_2$ level that the generator appears at. We find the generators ${K^a}_b$ at level $(0,0)$, corresponding to the $A_{n-2}$ sub-algebra of the gravity line, and the following other generators up to level $(1,1)$
\begin{alignat}{3}
\nonumber l_1 \longrightarrow \qquad &\qquad &\qquad &0 &\qquad &1\\ 
\nonumber l_2 \downarrow \qquad \qquad &0 &\qquad & R_0 & \qquad & R_1^{a_1a_2\ldots a_{(n-5)}}\\ 
&1 &\qquad &R_0^{ab} & \qquad &R_1^{a_1a_2\ldots a_{(n-4)},b}\label{Dgenerators}\\
\nonumber & & & & &R_1^{a_1a_2\ldots a_{(n-3)}}
\end{alignat}
The so-called dual to gravity $R_1^{a_1a_2\ldots a_{(n-4)},b}$ is listed amongst our low-level generators for completeness but it is not as well understood as the other listed generators and we will not consider it here, nor throughout this paper in any of the $\cal G^{+++}$ theories that we consider, as a starting point generator for finding the encoded brane solutions via (\ref{groupelement}). Each of the generators is associated with a root, $\beta$, such that $(\beta,\beta)=2$ except for the dilaton generator $R_0$ and the generator $R_1^{a_1a_2\ldots a_{(n-3)}}$, for which $(\beta,\beta)=0$. That $(\beta,\beta)=0$ for these generators means that we cannot commence with the generators $R_0$ and $R_1^{a_1a_2\ldots a_{(n-3)}}$ and use the group element in equation (\ref{groupelement}) to deduce an electric brane associated with them, as we would have a singularity coming from the $\frac{1}{(\beta,\beta)}$ for these generators. As such they are discarded as starting points for our method, as are all such generators which possess such a singularity in equation (\ref{groupelement}).\\

We make use of the following commutators, where we have chosen the coefficient $\frac{24}{D-2}$ in keeping with \cite{West1} and the other commutator has been determined from the Serre relations (\ref{Serre})
\begin{align}
\nonumber [R_0,R_0^{ab}]={\frac{24}{D-2}}R_0^{ab}\\
[R_0,R_1^{a_1a_2\ldots a_{n-5}}]=&-{\frac{24}{D-2}}R_1^{a_1a_2\ldots a_{n-5}}
\label{Dcommutators}
\end{align}
The simple root generators of $D_{n-3}^{+++}$ are
\begin{equation}
E_a={K^a}_{a+1}, a=1,\ldots (n-2), E_{(n-1)}=R_0^{(n-2)(n-1)}, E_n=R_1^{56\ldots(n-1)}
\end{equation}
and the Cartan sub-algebra generators, $H_a$, are given by
\begin{eqnarray}
\nonumber &H_a={K^a}_a-{K^{a+1}}_{a+1}, a=1,\ldots(n-2),\\
\nonumber &H_{(n-1)}=-\frac{2}{D-2}({K^1}_1+\ldots{K^{n-3}}_{n-3})+\frac{D-4}{D-2}({K^{n-2}}_{n-2}+{K^{n-1}}_{n-1})+{\frac{1}{6}}R_0 \\
&H_n=-\frac{D-4}{D-2}({K^1}_1+\ldots{K^4}_4)+\frac{2}{D-2}({K^5}_5+\ldots{K^{n-1}}_{n-1})-{\frac{1}{6}}R_0 \label{Dcartansub-algebra}
\end{eqnarray}
The low-level field content \cite{KleinschmidtSchnakenburgWest} is ${\hat{h}}_a\hspace{0pt}^b$, $A$, $A_{a_1a_2}$ and their field strengths have duals derived from  $A_{a_1\ldots a_{n-4},b}$, $A_{a_1\ldots a_{n-3}}$, $A_{a_1\ldots a_{n-5}}$ respectively. Our choice of local sub-algebra for the non-linear realisation allows us to write the group element as
\begin{align}
\nonumber g=\exp(\sum_{a\leq b}{\hat{h}}_a\hspace{0pt}^b{K^a}_b)&\exp(\frac{1}{(n-3)!}A_{a_1a_2\ldots a_{n-3}}R_1^{a_1a_2\ldots a_{n-3}})\\
&\exp(\frac{1}{(n-4)!}A_{a_1a_2\ldots a_{n-4},b}R_1^{a_1a_2\ldots a_{n-4},b}) \label{Dgroupelement}\\
\nonumber &\exp(\frac{1}{(n-5)!}A_{a_1a_2\ldots a_{n-5}}R_1^{a_1a_2\ldots a_{n-5}})\\
\nonumber &\exp(\frac{1}{2!}A_{ab}R_0^{ab})\exp(AR_0)
\end{align}
The field content of the associated maximally oxidised theory given in \cite{CremmerJuliaLuPope} agrees with the low-order $D_{n-3}^{+++}$ content given above. The Lagrangian for the oxidised theory, contains a graviton, $\phi$, a dilaton, $A$, and a 3-form field strength, $F_{\mu\nu\rho}=3\partial_{[\mu}A_{\nu\rho]}$, and is given by
\begin{equation}
A=\frac{1}{16\pi G_{n-1}}\int d^{n-1}x\sqrt{-g}(R-\frac{1}{2}\partial_\mu\phi\partial^\mu\phi-\frac{1}{2.3!}e^{\sqrt{\frac{8}{D-2}}\phi}{F_{\mu\nu\rho}}{F^{\mu\nu\rho}})
\label{Dlagrangian}
\end{equation}
We now demonstrate the use of the group element given in equation (\ref{groupelement}) in finding the electric branes of $D_{n-3}^{+++}$ from the generators listed in equation (\ref{Dgenerators}). We find the following electric branes
\paragraph{A String}
We commence by setting $\beta$ in (\ref{groupelement}) to be the root associated with the generator, $R_0^{(n-2)(n-1)}$ $(0 0\ldots 0 1 0)$, the highest weight in the $R^{a_1a_2}$ representation, corresponding to the field $A_{a1a2}$ in \cite{KleinschmidtSchnakenburgWest}. Our motivation is to find the usual BPS electric branes that emerge using the group element given. As such we wish to work with representations that have a time-like index, which we may obtain from our highest weight via multiple commutation with the appropriate ${K^a}_b$ generators such that all the indices on our operator are lowered to the lowest consecutive sequence of indices, including the time-like index. This has the consequence of finding the lowest weight generator in a particular representation. Our corresponding lowest weight in this representation is $R_0^{12} (1 2\ldots 2 1 1 0)$. The Cartan sub-algebra element associated with this lowest weight is found by using the root coefficients of the lowest weight, $(1 2\ldots 2 1 1 0)$, as coefficients of the Cartan sub-algebra elements associated with each root in the expansion of $\beta \cdot H$ so that
\begin{align}
\nonumber \beta \cdot H=&H_1+2(H_2+\ldots H_{n-3})+H_{n-2}+H_{n-1}\\
\nonumber =&({K^1}_1-{K^2}_2)+2({K^2}_2-{K^{n-2}}_{n-2})+({K^{n-2}}_{n-2}-{K^{n-1}}_{n-1})\\
\nonumber &-\frac{2}{D-2}({K^1}_1+\ldots{K^{n-3}}_{n-3})+\frac{D-4}{D-2}({K^{n-2}}_{n-2}+{K^{n-1}}_{n-1})\\
\nonumber &+{\frac{1}{6}}R_0\\
=&\frac{D-4}{D-2}({K^1}_1+{K^2}_2)-\frac{2}{D-2}({K^3}_3+\ldots{K^{n-1}}_{n-1})+{\frac{1}{6}}R_0 
\label{Dstringexpansion}
\end{align}
We now substitute equation (\ref{Dstringexpansion}) into our group element given in equation (\ref{groupelement}), noting that as $D_{n-3}^{+++}$ is a simply laced group $(\beta,\beta)=2$. The corresponding group element is
\begin{align}
\nonumber g_A=&\exp(-\frac{1}{2}\ln N_1(\frac{D-4}{D-2}({K^1}_1+{K^2}_2)-\frac{2}{D-2}({K^3}_3+\ldots{K^{n-1}}_{n-1})\\
\nonumber &+{\frac{1}{6}}R_0))\exp((1-N_1)E_{\beta})\\
\nonumber =&\exp(-\frac{1}{2}\ln N_1(\frac{D-4}{D-2}({K^1}_1+{K^2}_2)-\frac{2}{D-2}({K^3}_3+\ldots{K^{n-1}}_{n-1})))\\
&\exp(N_1^{-\frac{2}{D-2}}(1-N_1)E_{\beta})\exp(-{\frac{1}{12}}\ln N_1R_0) 
\label{Dstringelement}
\end{align}
Where we have moved the generator associated with the dilaton, $R_0$, to the right so that it agrees with the structure of the group element from which the non-linear realisation is constructed in equation (\ref{Dgroupelement}). We make use of $[R_0,R_0^{ab}]$ in equation (\ref{Dcommutators}) to do this and by examining the resulting $R_0$ term we find a dilaton which is given by
\begin{equation}
e^A=N_1^{-\frac{1}{12}}
\end{equation}
By reading off the coefficients of the ${K^a}_a$ in the group element we find a line element corresponding to a string
\begin{equation}
ds^2=N_1^{-\frac{D-4}{D-2}}(-dy_1^2+dx_2^2)+N_1^{\frac{2}{D-2}}(dx_3^2+\ldots+dx_{n-1}^2)
\label{Dstringlineelement}
\end{equation}
The brane is derived from a gauge potential which we can also read off from equation (\ref{Dstringelement}) as the coefficient of $E_\beta$
\begin{equation}
{A_{12}^T}_{(0)}=N_1^{-\frac{2}{D-2}}(1-N_1)
\end{equation}
This has tangent space indices, which we indicate with a $T$. To make the change to world volume indices we use the vierbein which we deduce from equation (\ref{Dstringlineelement}), $({e^{\hat{h}})^1}_1=({e^{\hat{h}})^2}_2=N_1^{-\frac{D-4}{2(D-2)}}$ and find
\begin{equation}
{A_{12}}_{(0)}=N_1^{-1}-1
\end{equation}
This gives rise to a 3-form field strength, which we label ${F_{\mu\nu\rho}}_0$.
\paragraph{An $(n-6)$ Brane}
Let us consider the representation $R^{a_1a_2\ldots a_{n-5}}$, corresponding to the field $A_{a_1a_2\ldots a_{n-5}}$ in \cite{KleinschmidtSchnakenburgWest} whose highest weight generator is $R_1^{5\ldots(n-1)}$ $(0 0\ldots 0 1)$. We take $\beta$ in equation (\ref{groupelement}) to be the root associated with the highest weight generator. The lowest weight generator in the representation is $R_1^{1\ldots(n-5)}$ $(1 2 3 4\ldots 4 3 2 1 0 1)$ and the corresponding element in the Cartan sub-algebra is
\begin{align}
\nonumber \beta \cdot H=&H_1+2H_2+3H_3+4(H_4+\ldots H_{n-5})+3H_{n-4}+2H_{n-3}+H_{n-2}+H_n\\
\nonumber =&{K^1}_1+{K^2}_2+{K^3}_3+{K^4}_4-{K^{n-4}}_{n-4}-{K^{n-3}}_{n-3}-{K^{n-2}}_{n-2}\\
\nonumber &-{K^{n-1}}_{n-1}-\frac{D-4}{D-2}({K^1}_1+\ldots{K^4}_4)+\frac{2}{D-2}({K^5}_5+\ldots{K^{n-1}}_{n-1})\\
\nonumber &-{\frac{1}{6}}R_0\\
=&\frac{2}{D-2}({K^1}_1+\ldots{K^{n-5}}_{n-5})-\frac{D-4}{D-2}({K^{n-4}}_{n-4}+\ldots{K^{n-1}}_{n-1})\\
\nonumber &-{\frac{1}{6}}R_0
\end{align}
Substituting this into equation (\ref{groupelement}), recalling that $(\beta, \beta)=2$, we find the corresponding group element is
\begin{align}
\nonumber g_A=&\exp(-\frac{1}{2}\ln N_{n-6}(\frac{2}{D-2}({K^1}_1+\ldots{K^{n-5}}_{n-5})\\
\nonumber &-\frac{D-4}{D-2}({K^{n-4}}_{n-4}+\ldots{K^{n-1}}_{n-1})-{\frac{1}{6}}R_0))\exp((1-N_{n-6})E_{\beta})\\
g_A=&\exp(-\frac{1}{2}\ln N_{n-6}(\frac{2}{D-2}({K^1}_1+\ldots{K^{n-5}}_{n-5})\\
\nonumber &-\frac{D-4}{D-2}({K^{n-4}}_{n-4}+\ldots{K^{n-1}}_{n-1})))\exp(N_{n-6}^{-\frac{2}{D-2}}(1-N_{n-6})E_{\beta})\\
\nonumber &\exp({\frac{1}{12}}\ln N_{n-6}R_0)
\end{align}
Where in the last line we have made use of $[R_0,R_1^{a_1\ldots a_{n-5}}]$, given in equation (\ref{Dcommutators}) to move the dilaton generator, $R_0$, to the far right of the expression in agreement with the group element from which the non-linear realisation is constructed (\ref{Dgroupelement}). By examining the $R_0$ term we find a dilaton given by
\begin{equation}
e^A=N_{n-6}^{\frac{1}{12}}
\end{equation}
And a line element corresponding to an electric $(n-6)$-brane
\begin{equation}
ds^2=N_{n-6}^{-\frac{2}{D-2}}(-dy_1^2+\ldots dx_{n-5}^2)+N_{n-6}^{\frac{D-4}{D-2}}(dx_{n-4}^2+\ldots+dx_{n-1}^2)
\end{equation}
The brane is derived from a gauge potential
\begin{equation}
{A_{12\ldots (n-5)}^T}_{(1)}=N_{n-6}^{-\frac{2}{D-2}}(1-N_{n-6})
\end{equation}
We use $({e^{\hat{h}})^1}_1=\ldots ({e^{\hat{h}})^{n-5}}_{n-5}=N_{n-6}^{-\frac{1}{D-2}}$ and make the change to world volume indices
\begin{align}
\nonumber {A_{12\ldots (n-5)}}_{(1)}=&N_{n-6}^{-\frac{2}{D-2}}N_{n-6}^{-\frac{n-5}{D-2}}(1-N_{n-6})\\
=&N_{n-6}^{-1}-1
\end{align}
Where we have used $D=n-1$ to get to the second line. This gauge potential is associated with an $(n-4)$-form field strength which we conclude is the dual of ${F_{\mu\nu\rho}}_0$.\\

We have have successfully reproduced all the usual BPS electric branes using the group element given in equation (\ref{groupelement}). The formulae we have found above do indeed correspond to the solutions of the Lagrangian (\ref{Dlagrangian}).\\

Let us now consider a fully worked example of the method outlined in section 1 to find a relation between our dilaton generator $A$ and $\phi$. We have found the gauge potentials $A^{(0)}, A_{a_1a_2}^{(0)}$ and $A_{a_1\ldots a_{n-5}}^{(1)}$ and we have one 3-form field strength appearing in the Lagrangian of equation (\ref{Dlagrangian}). Using equation (\ref{generalcovariantfieldstrength}) we write down the most general equation for the 3-form field strength
\begin{equation}
{F_{a_1a_2a_3}}_{(0)}=3e^{-c_{0,2}^{0,0}A}\partial_{[a_1}{A_{a_2a_3]}}_{(0)} \label{D3form}
\end{equation}

Its dual is an $(n-4)$-form field strength which is most generally
\begin{equation}
{F_{a_1\ldots a_{n-4}}}_{(1)}=(n-4)e^{-c_{0,n-5}^{0,1}A}\partial_{[a_1}{A_{a_2\ldots a_{n-5}]}}_{(1)} \label{D(n-4)form}
\end{equation}

These are related by duality giving
\begin{equation}
*{F_{a_1a_2a_3}}_{(0)}=\frac{1}{(n-4)!}\epsilon_{a_1a_2a_3b_1\ldots b_{n-4}}{F^{a_1\ldots a_3}}_{(0)}={F_{b_1\ldots b_{n-4}}}_{(1)} \label{Ddualityeq}
\end{equation}

Substituting equations (\ref{D3form}) and (\ref{D(n-4)form}) into equation (\ref{Ddualityeq}) yields
\begin{equation}
*(3e^{-c_{0,2}^{0,0}A}\partial_{[a_1}{A_{a_2a_3]}}_{(0)})=(n-4)e^{-c_{0,n-5}^{0,1}A}\partial_{[a_1}{A_{a_2\ldots a_{n-4}]}}_{(1)}
\end{equation}

If we bring the exponentials of $A$ together on the left-hand-side with the 3-form and differentiate we obtain
\begin{equation}
\partial_{[a_m}*(3e^{(-c_{0,2}^{0,0}+c_{0,n-5}^{0,1})A}\partial_{a_1}{A_{a_2a_3]}}_{(0)})=0 \label{Dfieldeq}
\end{equation}

Where the right-hand-side has vanished under differentiation by the Bianchi identity. The commutation of $H_{n-1}$ given in equation (\ref{Dcartansub-algebra}) with the positive root generators $E_n$ and $E_{n-1}$ enables us to find a relation between $c_{0,n-5}^{0,1}$ and $c_{0,2}^{0,0}$, we take $l={\frac{1}{12}}$ and we find
\begin{align}
\nonumber[H_{n-1},E_n]=&[H_{n-1},R^{56\ldots (n-1)}_1]=0\\
\nonumber & \Rightarrow l c_{0,n-5}^{0,1}=-\frac{4}{D-2}\\
\nonumber[H_{n-1},E_{n-1}]=&[H_{n-1},R^{(n-2)(n-1)}_0]=2R^{(n-2)(n-1)}_0\\
\nonumber & \Rightarrow l c_{0,2}^{0,0}=\frac{4}{D-2}\\
\Rightarrow \qquad & c_{0,2}^{0,0}=-c_{0,n-5}^{0,1}
\end{align}

Substituting this identity into our field equation (\ref{Dfieldeq}) gives
\begin{equation}
\partial_{[a_m}*(3e^{(-2c_{0,2}^{0,0})A}\partial_{a_1}{A_{a_2a_3]}}_{(0)})=0
\end{equation}

We have fixed our structure constants such that $c_{0,2}^{0,0}=\frac{24}{D-2}$ in accordance with reference \cite{West1} and by comparison with (\ref{Dlagrangian}) we find that
\begin{equation}
A=-\frac{1}{12}\sqrt{\frac{D-2}{2}}\phi
\end{equation}

We apply the method to $E_6^{+++}$ and $E_7^{+++}$ in Appendix A and show similarly that the low order field content derived from the group element (\ref{groupelement}) matches the field content given in reference \cite{CremmerJuliaLuPope} for the corresponding oxidised theories.
\section{Non-Simply Laced Groups}
The general method is slightly less straightforward for non-simply laced groups in that it becomes possible for the roots we find from the group theory to have a norm squared that differs from two. This difference manifests itself in two places. The first arises when we find the element in the Cartan sub-algebra corresponding to the lowest weight in a representation, at this point we must be wary of equation (\ref{Cartanbasis}) which indicates that we don't have an identity between the root coefficients and the Cartan sub-algebra coefficients but a proportionality factor of $\frac{(\alpha, \alpha)}{2}$. So that, for example, an element of the Cartan sub-algebra corresponding to a root with norm squared half that of the gravity line has an additional factor of $\frac{1}{2}$ in front of it. The second place we must be wary of the varying size of the simple roots comes when we make use of the group element where we must remember to put in the correct value of $(\beta, \beta)$ for the generator we are considering.
\subsection{Very Extended $B_{n-3}$}
The Dynkin diagram for $B_{n-3}^{+++}$ is shown in Appendix B, where the solid nodes indicate the gravity line we shall consider, which in this case is an $A_{n-2}$ sub-algebra.\\

We decompose the $B_{n-3}^{+++}$ algebra with respect to its $A_{n-2}$ sub-algebra by deleting the nodes corresponding to the simple roots $\alpha_n$ and $\alpha_{n-1}$ which have generators $R_1^{56\ldots(n-1)}(00\ldots 01)$ and $R_0^{(n-1)}(00\ldots 10)$ respectively. We associate the levels $l_1$ and $l_2$ with these respectively and the $s$ labels on our generators are chosen to be the $l_1$ level of that generator. From reference \cite{KleinschmidtSchnakenburgWest} we find the $B_{n-3}^{+++}$ algebra contains the generators ${K^a}_b$ at level $(0,0)$, corresponding to the $A_{n-2}$ sub-algebra of the gravity line. We have the following other generators up to level $(1,2)$ \cite{KleinschmidtSchnakenburgWest}
\begin{alignat}{3}
\nonumber l_1 \longrightarrow \qquad &\qquad &\qquad &0 &\qquad &1\\
\nonumber l_2 \downarrow \qquad \qquad &0 & & R_0 & &R_1^{a_1a_2\ldots a_{(n-5)}}\\
&1 & &R_0^{a} & &R_1^{a_1a_2\ldots a_{(n-4)}} \label{Bgenerators}\\
\nonumber &2 & &R_0^{a_1a_2} & &R_1^{a_1a_2\ldots a_{(n-4)},b}\\
\nonumber & & & & &R_1^{a_1a_2\ldots a_{(n-3)}}
\end{alignat}
The generators $R_0^{a}$ and $R_1^{a_1a_2\ldots a_{n-4}}$ have associated roots $\beta$ such that $(\beta,\beta)=1$, the rest and the gravity line nodes have $(\beta,\beta)=2$, with the exception of the dilaton generator $R_0$ and the generator $R_1^{a_1a_2\ldots a_{(n-3)}}$ which have $(\beta,\beta)=0$. We cannot apply our method to $R_0$ and $R_1^{a_1a_2\ldots a_{(n-3)}}$ so they are discarded as starting points when we come to find the electric branes from the generators.\\

We make use of the following commutator relations where we have chosen the commutator coefficient of $[R_0,R_0^{a_1}]$ and the rest have followed from the relations (\ref{jicoefficients1})-(\ref{jicoefficients4}) and the Serre relations (\ref{Serre})
\begin{align}
\nonumber &[R_0,R_0^{a_1}]=\frac{1}{4}\sqrt{\frac{8}{D-2}}R_0^{a_1}\\
&[R_0,R_0^{a_1a_2}]=\frac{1}{2}\sqrt{\frac{8}{D-2}}R_0^{a_1a_2}\label{Bcommutators}\\
\nonumber &[R_0,R_1^{a_1a_2\ldots a_{{n-5}}}]=-\frac{1}{2}\sqrt{\frac{8}{D-2}}R_1^{a_1a_2\ldots a_{n-5}}\\
\nonumber &[R_0,R_1^{a_1a_2\ldots a_{{n-4}}}]=-\frac{1}{4}\sqrt{\frac{8}{D-2}}R_1^{a_1a_2\ldots a_{n-4}}
\end{align}
We note from content of the table given in \cite{KleinschmidtSchnakenburgWest} for $B_{n-3}^{+++}$ that no higher order commutators are needed.\\

The simple root generators of $B_{n-3}^{+++}$ are
\begin{equation}
E_a={K^a}_{a+1}, a=1,\ldots (n-2), E_{(n-1)}=R_0^{(n-1)}, E_n=R_1^{56\ldots(n-1)}
\end{equation}
and the Cartan sub-algebra generators, $H_a$ are given by
\begin{eqnarray}
\nonumber &H_a={K^a}_a-{K^{a+1}}_{a+1}, a=1,\ldots(n-2),\\
\nonumber &H_{(n-1)}=-\frac{2}{D-2}({K^1}_1+\ldots{K^{n-2}}_{n-2})+2\frac{D-3}{D-2}({K^{n-1}}_{n-1})\\
&+\sqrt{\frac{8}{D-2}}R_0,\\
\nonumber &H_n=-\frac{D-4}{D-2}({K^1}_1+\ldots{K^4}_4)+\frac{2}{D-2}({K^5}_5+\ldots{K^{n-1}}_{n-1})\\
\nonumber &-\sqrt{\frac{8}{D-2}}R_0
\end{eqnarray}
The low-level field content \cite{KleinschmidtSchnakenburgWest} is ${\hat{h}}_a\hspace{0pt}^b$, $A$, $A_{a_1}$, $A_{a_1a_2}$ and their field strengths have duals derived from  $A_{a_1\ldots a_{n-4},b}$, $A_{a_1\ldots a_{n-3}}$, $A_{a_1\ldots a_{n-4}}$, $A_{a_1\ldots a_{n-5}}$ respectively. Our choice of local sub-algebra for the non-linear realisation allows us to write the group element of $B_{n-3}^{+++}$ as
\begin{align}
\nonumber g=\exp(\sum_{a\leq b}{\hat{h}}_a\hspace{0pt}^b{K^a}_b)&\exp(\frac{1}{(n-3)!}A_{a_1a_2\ldots a_{n-3}}R_1^{a_1a_2\ldots a_{n-3}})\\
\nonumber &\exp(\frac{1}{(n-4)!}A_{a_1a_2\ldots a_{n-4},b}R_1^{a_1a_2\ldots a_{n-4},b})\\
&\exp(\frac{1}{(n-4)!}A_{a_1a_2\ldots a_{n-4}}R_1^{a_1a_2\ldots a_{n-4}})\\
\nonumber &\exp(\frac{1}{(n-5)!}A_{a_1a_2\ldots a_{n-5}}R_1^{a_1a_2\ldots a_{n-5}})\\
\nonumber &\exp(\frac{1}{2!}A_{ab}R_0^{ab})\exp(A_aR_0^a)\exp(AR_0)
\end{align}
The field content of the associated maximally oxidised theory given in \cite{CremmerJuliaLuPope} agrees with the low-order $B_{n-3}^{+++}$ content given above. The Lagrangian for the oxidised theory, containing a graviton, $\phi$, a 2-form field strength, ${F_{\mu\nu}}_0=2\partial_{[\mu}{A_{\nu]}}_0$, a 3-form field strength, ${F_{\mu\nu\rho}}_0=3(\partial_{[\mu}{A_{\nu\rho]}}_0+{A_{[\mu}}_0\partial_{\nu}{A_{\rho]}}_0)$ and a dilaton, $A$, is
\begin{align}
A=\frac{1}{16\pi G_{n-1}}\int d^{n-1}x\sqrt{-g}(&R-\frac{1}{2}\partial_\mu\phi\partial^\mu\phi-\frac{1}{2.2!}e^{\frac{1}{2}\sqrt{\frac{8}{D-2}}\phi}{{F_{\mu\nu}}_0}{F^{\mu\nu}_0}\label{Blagrangian}\\
\nonumber &-\frac{1}{2.3!}e^{\sqrt{\frac{8}{D-2}}\phi}{F_{\mu\nu\rho}}_0F^{\mu\nu\rho}_0)
\end{align}
We now demonstrate the use of the group element (\ref{groupelement}) to generate all of the usual BPS electric branes of $B_{n-3}^{+++}$ starting from the generators given in equation (\ref{Bgenerators}). We find the following electric branes
\paragraph{A Particle}
We wish to find the electric brane associated with the representation $R_0^a$ that has highest weight generator $R_0^{(n-1)} (0 0\ldots 0 1 0)$, and so we set $\beta$ in equation $\ref{groupelement}$ to be its corresponding root. We note that $(\beta,\beta)=1$. The lowest weight generator in the representation is $R_0^{1} (1 1\ldots 1 1 0)$. To find the corresponding element in the Cartan sub-algebra we observe from Appendix B that only the root corresponding to the $(n-1)$th node does not have norm squared equal to two. Indeed $(\alpha_{n-1},\alpha_{n-1})=1$, so from equation (\ref{Cartanbasis}) we see that we have to effectively halve the $H_{n-1}$ generator contribution when we find $\beta\cdot H$ from the root coefficients, $\alpha^i$, so that
\begin{align}
\nonumber \beta \cdot H=&H_1+\ldots +H_{n-2}+\frac{1}{2}H_{n-1}\\
=&\frac{D-3}{D-2}({K^1}_1)-\frac{1}{D-2}({K^2}_2+\ldots+{K^{n-1}}_{n-1})+\sqrt{\frac{2}{D-2}}R_0
\label{Bparticleexpansion}
\end{align}
By substituting (\ref{Bparticleexpansion}) into equation (\ref{groupelement}), and bearing in mind that $(\beta,\beta)=1$ for $R_0^1$, we find the corresponding group element is
\begin{align}
\nonumber g_A=&\exp(-\ln N_0(\frac{D-3}{D-2}({K^1}_1)-\frac{1}{D-2}({K^2}_2+\ldots+{K^{n-1}}_{n-1})\\
\nonumber &+\sqrt{\frac{2}{D-2}}R_0))\exp((1-N_0)E_{\beta})\\
=&\exp(-\ln N_0(\frac{D-3}{D-2}({K^1}_1)-\frac{1}{D-2}({K^2}_2+\ldots+{K^{n-1}}_{n-1})))\\
\nonumber &\exp(N_0^{-\frac{1}{D-2}}(1-N_0)E_{\beta})\exp(-\sqrt{\frac{2}{D-2}}\ln N_0R_0) 
\end{align}
In the last line we have made use of $[R_0, R_0^{a_1}]$ from equation (\ref{Bcommutators}) to move the dilaton generator to the far right so that it agrees with the group element from which the non-linear realisation is constructed. By examining the $R_0$ term we find a dilaton given by
\begin{equation}
e^A=N_0^{-\sqrt{\frac{2}{D-2}}}
\end{equation}
And a line element corresponding to a particle
\begin{equation}
ds^2=N_0^{-\frac{2(D-3)}{D-2}}(-dy_1^2)+N_0^{\frac{2}{D-2}}(dx_2^2+\ldots+dx_{n-1}^2)
\end{equation}
The particle is derived from a gauge potential
\begin{equation}
{A_{1}^T}_{(0)}=N_0^{-\frac{1}{D-2}}(1-N_0)
\end{equation}
We complete the change to world volume indices using $({e^{\hat{h}})^1}_1=N_0^{-\frac{D-3}{D-2}}$
\begin{equation}
{A_{1}}_{(0)}=N_1^{-1}-1
\end{equation}
This gives rise to a 2-form field strength, which we label ${F_{\mu\nu}}_0$.
\paragraph{An $(n-5)$ Brane}
Proceeding in the usual manner we find the electric brane associated with the representation $R_1^{a_1a_2\ldots a_(n-4)}$whose highest weight generator is $R_1^{45\ldots (n-1)}$ $(0001\ldots 111)$, whose associated root we now take as $\beta$. The lowest weight generator is $R_1^{12\ldots (n-4)}$ $(1234\ldots 43211)$ and the corresponding element in the Cartan sub-algebra is, recalling that the root associated with the $(n-1)$th node of the Dynkin diagram is short, being half the length of the other roots,
\begin{align}
\nonumber \beta \cdot H=&H_1+2H_2+3H_3+4(H_4+\ldots +H_{n-4})\\
\nonumber &+3H_{n-3}+2H_{n-2}+\frac{1}{2}H_{n-1}+H_n\\
=&\frac{1}{D-2}({K^1}_1+\ldots {K^{n-4}}_{n-4})-\frac{D-3}{D-2}({K^{n-3}}_{n-3}+\ldots{K^{n-1}}_{n-1})\\
\nonumber &-\frac{1}{2}\sqrt{\frac{8}{D-2}}R_0
\end{align}
We note that $(\beta,\beta)=1$ and substitute this expression into equation (\ref{groupelement}) to find that the group element for the generator $R_1^{45\ldots (n-1)}$ is
\begin{align}
\nonumber g_A=\exp(-\ln N_{n-5}&(\frac{1}{D-2}({K^1}_1+\ldots {K^{n-4}}_{n-4})\\
&-\frac{D-3}{D-2}({K^{n-3}}_{n-3}+\ldots{K^{n-1}}_{n-1})))\\
\nonumber &\exp(N_{n-5}^{-\frac{1}{D-2}}(1-N_{n-5})E_{\beta})\\
\nonumber &\exp(\sqrt{\frac{2}{D-2}}\ln N_{n-5}R_0) 
\end{align}
We find a dilaton given by
\begin{equation}
e^A=N_{n-5}^{\sqrt{\frac{2}{D-2}}}
\end{equation}
And a line element corresponding to an $(n-5)$-brane
\begin{equation}
ds^2=N_{n-5}^{-\frac{2}{D-2}}(-dy_1^2+dx_2^2+\ldots dx_{n-4}^2)+N_{n-5}^{\frac{2(D-3)}{D-2}}(dx_{n-3}^2+\ldots+dx_{n-1}^2)
\end{equation}
The particle is derived from a gauge potential
\begin{equation}
{A_{12\ldots (n-4)}^T}_{(1)}=N_{n-5}^{-\frac{1}{D-2}}(1-N_{n-5})
\end{equation}
We use $({e^{\hat{h}})^1}_1=\ldots ({e^{\hat{h}})^{n-4}}_{n-4}= N_{n-5}^{\frac{-1}{D-2}}$ to complete the change to world volume indices
\begin{equation}
{A_{12\ldots (n-4)}}_{(1)}=N_{n-5}^{-1}-1
\end{equation}
This gives rise to an $(n-3)$ form field strength, which we interpret to be the dual of ${F_{\mu\nu}}_0$.
\paragraph{A String}
We wish to find the electric brane associated with the representation $R_0^{a_1a_2}$ with the highest weight generator $R_0^{(n-2)(n-1)} (0 0\ldots 0 1 2 0)$, whose corresponding root we take to be $\beta$ and we note that $(\beta,\beta)=2$. The lowest weight generator is $R_0^{12} (1 2\ldots 2 2 0)$ and the corresponding element in the Cartan sub-algebra is
\begin{align}
\nonumber \beta \cdot H=&H_1+2(H_2+\ldots +H_{n-2})+H_{n-1}\\
=&\frac{D-4}{D-2}({K^1}_1+{K^2}_2)-\frac{2}{D-2}({K^3}_3+\ldots {K^{n-1}}_{n-1})+\sqrt{\frac{8}{D-2}}R_0
\end{align}
The corresponding group element is
\begin{align}
\nonumber g_A=&\exp(-\frac{1}{2}\ln N_1(\frac{D-4}{D-2}({K^1}_1+{K^2}_2)-\frac{2}{D-2}({K^3}_3+\ldots {K^{n-1}}_{n-1})))\\
&\exp(N_1^{-\frac{2}{D-2}}(1-N_1)E_{\beta})\exp(-\sqrt{\frac{2}{D-2}}\ln N_1R_0) 
\end{align}
We find a dilaton given by
\begin{equation}
e^A=N_0^{-\sqrt{\frac{2}{D-2}}}
\end{equation}
And a line element corresponding to a string
\begin{equation}
ds^2=N_1^{-\frac{D-4}{D-2}}(-dy_1^2+dx_2^2)+N_1^{\frac{2}{D-2}}(dx_3^2+\ldots+dx_{n-1}^2)
\end{equation}
The string is derived from a gauge potential
\begin{equation}
{A_{12}^T}_{(0)}=N_1^{-\frac{2}{D-2}}(1-N_1)
\end{equation}
We complete the change to world volume indices using $({e^{\hat{h}})^1}_1=({e^{\hat{h}})^2}_2=N_1^{-\frac{D-4}{2(D-2)}}$
\begin{equation}
{A_{12}}_{(0)}=N_1^{-1}-1
\end{equation}
From this we derive a 3-form field strength, which we label ${F_{\mu\nu\rho}}_0$.
\paragraph{An $(n-6)$ Brane}
We wish to find the electric brane associated with the representation $R_1^{a_1a_2\ldots a_{n-5}}$ with highest weight generator $R_1^{56\ldots (n-1)}$ $(0 0 0\ldots 0 0 1)$, whose corresponding root we take as $\beta$ in equation (\ref{groupelement}) and we note that $(\beta, \beta)=2$. The associated lowest weight generator in this representation is $R_1^{12\ldots (n-5)}$ $(1 2 3 4\ldots 4 3 2 1 0 1)$ and the corresponding element in the Cartan sub-algebra is
\begin{align}
\nonumber \beta \cdot H=&H_1+2H_2+3H_3+4(H_4+\ldots +H_{n-5})\\
\nonumber &+3H_{n-4}+2H_{n-3}+H_{n-2}+H_n\\
=&\frac{2}{D-2}({K^1}_1+\ldots {K^{n-5}}_{n-5})-\frac{D-4}{D-2}({K^{n-4}}_{n-4}+\ldots{K^{n-1}}_{n-1})\\
\nonumber &-\sqrt{\frac{8}{D-2}}R_0
\end{align}
The corresponding group element is
\begin{align}
\nonumber g_A=\exp(-\frac{1}{2}\ln N_{n-6}(&\frac{2}{D-2}({K^1}_1+\ldots {K^{n-5}}_{n-5})\\
&-\frac{D-4}{D-2}({K^{n-4}}_{n-4}+\ldots{K^{n-1}}_{n-1})))\\
\nonumber &\exp(N_{n-6}^{-\frac{2}{D-2}}(1-N_{n-6})E_{\beta})\\
\nonumber &\exp(\sqrt{\frac{2}{D-2}}\ln N_{n-6}R_0) 
\end{align}
We find a dilaton given by
\begin{equation}
e^A=N_{n-6}^{\sqrt{\frac{2}{D-2}}}
\end{equation}
And a line element corresponding to an $(n-6)$-brane
\begin{equation}
ds^2=N_{n-6}^{-\frac{2}{D-2}}(-dy_1^2+dx_2^2+\ldots dx_{n-5}^2)+N_{n-6}^{\frac{D-4}{D-2}}(dx_{n-4}^2+\ldots+dx_{n-1}^2)
\end{equation}
The brane is derived from a gauge potential
\begin{equation}
{A_{12\ldots (n-5)}^T}_{(1)}=N_{n-6}^{-\frac{2}{D-2}}(1-N_{n-6})
\end{equation}
We use $({e^{\hat{h}})^1}_1=\ldots ({e^{\hat{h}})^{n-5}}_{n-5}= N_{n-6}^{-\frac{1}{D-2}}$ to complete the change to world volume indices
\begin{equation}
{A_{12\ldots (n-5)}}_{(1)}=N_{n-6}^{-1}-1
\end{equation}
This gives rise to an $(n-4)$-form field strength, which we interpret to be the dual of ${F_{\mu\nu\rho}}_0$.\\

We have have successfully reproduced all the usual BPS electric branes using the group element given in equation (\ref{groupelement}). The formulae we have found above do indeed correspond to the solutions of the Lagrangian (\ref{Blagrangian}).\\

We find our dilaton field $A$ to be related to $\phi$ by

\begin{equation}
A = -\phi
\end{equation}

\subsection{Very Extended $F_4$}
The ${F_4}^{+++}$ Dynkin diagram is shown in Appendix B, where the solid nodes represent the gravity line with respect to which we decompose our Kac-Moody algebra. The shorter roots, enumerated as nodes 6 and 7 have $(\alpha,\alpha)=1$ where the gravity line roots are normalised to have norm squared of two.\\

The $F_4^{+++}$ algebra is decomposed with respect to its $A_5$ sub-algebra, the simple roots whose nodes we delete are $\alpha_7$ and $\alpha_6$ and their generators are $R_1 (0000001)$ and $R_0^6(0000010)$. We associate the levels $l_1$ and $l_2$ with these respectively. The $s$ labels on our generators are chosen to be the $l_1$ level of the generator. From reference \cite{KleinschmidtSchnakenburgWest} we find the ${F_4}^{+++}$ algebra decomposed with respect to an $A_5$ sub-algebra contains the generators ${K^a}_b$ at level (0,0) and the following generators at levels $(l_1, l_2)$
\begin{alignat}{4}
\nonumber l_1 \longrightarrow \qquad &\qquad &\qquad &0 &\qquad &1 &\qquad &2 \\ 
\nonumber l_2 \downarrow \qquad \qquad &0 &\qquad &R_0 & \qquad &R_1 & \qquad & \\ 
\nonumber &1 &\qquad &R_0^a & \qquad & R_1^a & \qquad & \\ 
&2 &\qquad &R_0^{ab} & \qquad &R_1^{ab} & \qquad &{R_2}^{ab} \label{Fgenerators}\\ 
\nonumber &3 &\qquad & & \qquad &R_1^{abc} & \qquad &{R_2}^{abc} \\ 
\nonumber &4 &\qquad & & \qquad &R_1^{abcd} & \qquad &{R_2}^{abc,d}\\
\nonumber & &\qquad & & \qquad & & \qquad &{R_2}^{abcd} 
\end{alignat}
The generators $R_0$ and ${R_2}^{abcd}$ have associated roots $\beta$ such that$(\beta,\beta)=0$ so we discard them as starting points for our method in this section, and ${R_2}^{abc,d}$, ${R_2}^{ab}$, $R_0^{ab}$ have $(\beta,\beta)=2$, the other generators all have $(\beta,\beta)=1$.
We make use of the following commutation relations where we have chosen the commutator coefficient for $[R_0,R_0^a]$ and the rest have been deduced from equations (\ref{jicoefficients1})-(\ref{jicoefficients4}) and the Serre relations (\ref{Serre})
\begin{alignat}{3}
\nonumber &[R_0,R_0^a]= \frac{1}{\sqrt{8}}R_0^a &\qquad &[R_0,R_1^a]=-\frac{1}{\sqrt{8}}R_1^a &\qquad & \\
\nonumber &[R_0,R_0^{ab}]= \frac{1}{\sqrt{2}}R_0^{ab} &\qquad &[R_0,R_1^{ab}]=0 &\qquad &[R_0,{R_2}^{ab}]=-\frac{1}{\sqrt{2}}{R_2}^{ab} \label{Fcommutators}\\
& &\qquad &[R_0,R_1^{abc}]= \frac{1}{\sqrt{8}}R_1^{abc} &\qquad &[R_0,{R_2}^{abc}]=-\frac{1}{\sqrt{8}}{R_2}^{abc} \\
\nonumber & &\qquad &[R_0,R_1^{abcd}]= \frac{1}{\sqrt{2}}R_1^{abcd} &\qquad &[R_0,{R_2}^{abc,d}]=0 
\end{alignat}
The simple root generators of ${F_4}^{+++}$ are
\begin{equation}
E_a={K^a}_{a+1}, a=1,\ldots5, E_6=R_0^6, E_7=R_1
\end{equation}
and the Cartan sub-algebra generators, $H_a$,is given by \cite{EnglertHouart}
\begin{eqnarray}
\nonumber &H_a={K^a}_a-{K^{a+1}}_{a+1}, a=1,\ldots5,\\
&H_6=-\frac{1}{2}({K^1}_1+\ldots{K^5}_5)+\frac{3}{2}{K^6}_6+\sqrt{2}R_0,\\
\nonumber &H_7=-\sqrt{8}R_0
\end{eqnarray}
The low-level field content \cite{KleinschmidtSchnakenburgWest} is ${\hat{h}}_a\hspace{0pt}^b$, $A$, ${A_{a_1}}_0$, ${A_{a_1}}_1$, ${A_{a_1a_2}}_2$, ${A_{a_1a_2a_3a_4}}_1$ and their field strengths have duals derived from  ${A_{a_1a_2a_3,b}}_2$, ${A_{a_1a_2a_3a_4}}_2$ ${A_{a_1a_2a_3}}_2$, ${A_{a_1a_2a_3}}_1$, ${A_{a_1a_2}}_0$, $A_1$, respectively, we also have an ${A_{a_1a_2}}_1$ field without a dual listed above, so we treat it as a self-dual field. Our choice of local sub-algebra for the non-linear realisation allows us to write the group element of ${F_4}^{+++}$ as
\begin{align}
\nonumber g=\exp(\sum_{a\leq b}{\hat{h}}_a\hspace{0pt}^b{K^a}_b)&\exp(\frac{1}{4!}{A_{a_1a_2a_3a_4}}_1R_1^{a_1a_2a_3a_4})\exp(\frac{1}{4!}{A_{a_1a_2a_3a_4}}_2{R_2}^{a_1a_2a_3a_4})\\
\nonumber &\exp(\frac{1}{3!}{A_{a_1a_2a_3,b}}_2{R_2}^{a_1a_2a_3,b})\exp(\frac{1}{3!}{A_{a_1a_2a_3}}_2{R_2}^{a_1a_2a_3})\\
&\exp(\frac{1}{3!}{A_{a_1a_2a_3}}_1R_1^{a_1a_2a_3})\exp(\frac{1}{2!}{A_{a_1a_2}}_2{R_2}^{a_1a_2})\\
\nonumber &\exp(\frac{1}{2!}{{A_{a_1a_2}}_1R_1^{a_1a_2})\exp(\frac{1}{2!}A_{a_1a_2}}_0R_0^{a_1a_2})\\
\nonumber &\exp({A_{a_1}}_1R_1^a)\exp({A_{a_1}}_0R_0^{a_1})\exp(A_1R_1)\exp(AR_0)
\end{align}
The field content of the associated maximally oxidised theory given in \cite{CremmerJuliaLuPope} agrees with the low-order $F_4^{+++}$ content given above. The Lagrangian for the oxidised theory, contains two 2-form field strengths ${F_{\mu\nu}}_0=2\partial_{[\mu}{A_{\nu]}}_0+\sqrt{2}A_1\partial_{[\mu}{A_{\nu]}}_1$ and ${F_{\mu\nu}}_1=2\partial_{[\mu}{A_{\nu]}}_1$, a 3-form field strength ${F_{\mu\nu\rho}}_2=3(\partial_{[\mu}{A_{\nu\rho]}}_2+{A_{[\mu}}_1\partial_{\nu}{A_{\rho]}}_1)$, a self-dual 3-form field strength ${F_{\mu\nu\rho}}_1=3(\partial_{[\mu}{A_{\nu\rho]}}_1-{A_{[\mu}}_0\partial_{\nu}{A_{\rho]}}_1)-\frac{1}{\sqrt{2}}{A_1}{F_{\mu\nu\rho}}_2$, a 1-form field strength ${F_\mu}_1=\partial_{\mu}A_1$, and the dilaton $A$.
\begin{align}
\nonumber A=\frac{1}{16\pi G_6}\int d^6x\sqrt{-g}(&R-\frac{1}{2}\partial_\mu\phi\partial^\mu\phi-\frac{1}{2.2!}e^{{\frac{1}{\sqrt{2}}}\phi}{F_{\mu\nu}}_0F^{\mu\nu}_0\\
&-\frac{1}{2.2!}e^{{-\frac{1}{\sqrt{2}}}\phi}{F_{\mu\nu}}_1F^{\mu\nu}_1-\frac{1}{2.3!}e^{-\sqrt{2}\phi}{{F_{\mu\nu\rho}}_2}F^{\mu\nu\rho}_2\\
\nonumber &-\frac{1}{2}e^{\sqrt{2}\phi}{{F_{\mu}}_1}F^{\mu}_1)\\
\nonumber -\int_{C.S.} d^6x\epsilon^{a_1\ldots a_6}&(\frac{1}{\sqrt{2}.3!3!}A_1{F_{a_1a_2a_3}}_2{F_{a_4a_5a_6}}_1+\frac{1}{2.2!3!}{A_{a_1}}_0{F_{a_2a_3}}_0{F_{a_4a_5a_6}}_2\\
\nonumber &+\frac{1}{2.2!3!}{A_{a_1}}_0{F_{a_2a_3}}_1{F_{a_4a_5a_6}}_1)
\label{Flagrangian}
\end{align}
The reader must remember that ${F_{\mu\nu\rho}}_1$ is self-dual and its kinetic term in the action integral vanishes. We again use the group element (\ref{groupelement}) to generate all of the electric branes of $F_4^{+++}$ starting from the generators given in equation (\ref{Fgenerators}). We find the following electric branes
\paragraph{A Particle}
Commencing by setting $\beta$ to be the root associated with generator $R_0^6$ $(0 0 0 0 0 1 0)$, the highest weight of the ${R^{a_1}}_0$ representation. Noting that $(\beta,\beta)=1$ we proceed and find the lowest weight generator in this representation is $R_0^1$ $(1 1 1 1 1 1 0)$. Taking account of the shorter root on the sixth node of the ${F_4}^{+++}$ Dynkin diagram, we use equation (\ref{Cartanbasis}) to find the element corresponding to $R_0^1$ in the Cartan sub-algebra
\begin{align}
\nonumber \beta \cdot H&=(H_1+\ldots H_5)+\frac{1}{2}H_6\\
&=\frac{3}{4}({K^1}_1)-\frac{1}{4}({K^2}_2+\ldots{K^6}_6)+\frac{1}{\sqrt{2}}R_0
\end{align}
And from equation (\ref{groupelement}) we find the corresponding group element is
\begin{align}
\nonumber g_A =& \exp(-\ln N_0(\frac{3}{4}({K^1}_1)-\frac{1}{4}({K^2}_2+\ldots{K^6}_6)+\frac{1}{\sqrt{2}}R_0))\exp((1-N_0)E_\beta)\\
=& \exp(-\ln N_0(\frac{3}{4}({K^1}_1)-\frac{1}{4}({K^2}_2+\ldots{K^6}_6)))\exp(N_0^{-\frac{1}{4}}(1-N_0)E_\beta)\\
\nonumber &\exp(-\frac{1}{\sqrt{2}}\ln N_0R_0)
\end{align} 
So we find a dilaton given by 
\begin{equation}
e^A = N_0^{-\frac{1}{\sqrt{2}}}
\end{equation}
And a line element corresponding to a particle
\begin{equation}
ds^2=N_0^{-\frac{3}{2}}(-dy_1^2)+N_0^{\frac{1}{2}}(dx_2^2+\ldots+dx_6^2)
\label{Fparticlelinelement}
\end{equation}
We have a gauge field given by
\begin{equation}
{A_1^T}_{(0)} = N_0^{-\frac{1}{4}}(1-N_0)
\end{equation}
We change to world volume indices via the vielbein ${(e^h)^1}_1=N_0^{-\frac{3}{4}}$
\begin{equation}
{A_1}_{(0)}=N_0^{-1}-1
\end{equation}
We associate this gauge potential with a 2-form field strength, ${F_{\mu\nu}}_0$.
\paragraph{A 2-Brane}
We wish to find the electric brane associated with the representation ${R^{a_1a_2a_3}}_2$ whose highest weight generator is ${R_2}^{456} (0 0 0 1 2 3 2)$. We take this to be $\beta$ in equation (\ref{groupelement}) and we note that $(\beta,\beta)=1$. The lowest weight generator is ${R_2}^{123} (1 2 3 3 3 3 2)$ and the corresponding element in the Cartan sub-algebra is
\begin{align}
\nonumber \beta \cdot H=&H_1+2H_2+3(H_3+\ldots H_5)+\frac{3}{2}H_6+H_7\\
=&\frac{1}{4}({K^1}_1+\ldots{K^3}_3)-\frac{3}{4}({K^4}_4+\ldots{K^6}_6)-\frac{1}{\sqrt{2}}R_0
\end{align}
Substituting this into equation (\ref{groupelement}) we find the appropriately arranged element is
\begin{align}
\nonumber g_A=&\exp(-\ln N_2(\frac{1}{4}({K^1}_1+\ldots{K^3}_3)-\frac{3}{4}({K^4}_4+\ldots{K^6}_6)-\frac{1}{\sqrt{2}}\ln N_2R_0))\\
\nonumber &\exp((1-N_2)E_{\beta})\\
=&\exp(-\ln N_2(\frac{1}{4}({K^1}_1+\ldots{K^3}_3)-\frac{3}{4}({K^4}_4+\ldots{K^6}_6)\\
\nonumber &\exp(N_2^{-\frac{1}{4}}(1-N_2)E_{\beta})\exp(+\frac{1}{\sqrt{2}}\ln N_2R_0)
\end{align}
We find a dilaton given by
\begin{equation}
e^A=N_2^{+\frac{1}{\sqrt{2}}}
\end{equation}
And a line element corresponding to a 2-brane
\begin{equation}
ds^2=N_2^{-\frac{1}{2}}(-dy_1^2+\ldots dx_3^2)+N_2^{\frac{3}{2}}(dx_4^2+\ldots+dx_6^2) \label{F2branelineeelement}
\end{equation}
The brane is derived from the gauge field given by
\begin{equation}
{A_{123}^T}_{(2)}=N_2^{-\frac{1}{4}}(1-N_2)
\end{equation}
We complete the change to world volume indices using $({e^{\hat{h}})^1}_1=\ldots({e^{\hat{h}})^3}_3=N_0^{-\frac{1}{4}}$
\begin{equation}
{A_{123}}_{(2)}=N_2^{-1}-1
\end{equation}
We conclude that ${R_2}^{456}$ is the generator associated with the dual of ${F_{\mu\nu}}_0$.
\paragraph{A Second Particle}
The calculation for the representation ${R^{a_1}}_1$ which has a highest weight generator $R_1^6$ is the same as that for $R_0^6$ except that the expansion of $\beta \cdot H$ has an extra $\frac{1}{2}H_7$ added on. That is,
\begin{equation}
\beta \cdot H=\frac{3}{4}({K^1}_1)-\frac{1}{4}({K^2}_2+\ldots{K^6}_6)-\frac{1}{\sqrt{2}}R_0
\end{equation}
And from equation (\ref{groupelement}) we find the corresponding group element is
\begin{align}
\nonumber g_A =& \exp(-\ln N_0(\frac{3}{4}({K^1}_1)-\frac{1}{4}({K^2}_2+\ldots{K^6}_6)-\frac{1}{\sqrt{2}}R_0))\exp((1-N_0)E_\beta)\\
=& \exp(-\ln N_0(\frac{3}{4}({K^1}_1)-\frac{1}{4}({K^2}_2+\ldots{K^6}_6)))\exp(N_0^{-\frac{1}{4}}(1-N_0)E_\beta)\\
\nonumber &\exp(\frac{1}{\sqrt{2}}\ln N_0R_0)
\end{align} 
We find a line element identical to (\ref{Fparticlelinelement}) for a particle and a dilaton given by
\begin{equation}
e^A = N_0^{\frac{1}{\sqrt{2}}}
\end{equation}
We have a gauge field given by
\begin{equation}
{A_1^T}_{(1)} = N_0^{-\frac{1}{4}}(1-N_0)
\end{equation}
We change to world volume indices via the vielbein ${(e^h)^1}_1=N_0^{-\frac{3}{4}}$ and find
\begin{equation}
{A_1}_{(1)}=N_0^{-1}-1
\end{equation}
We associate this gauge potential with a 2-form field strength which we label ${F_{\mu\nu}}_1$.
\paragraph{A Second 2-Brane}
Similarly, the derivation of the electric brane associated with the highest weight generator $R_1^{456}$ is the same as that for ${R_2}^{456}$ but with a $\frac{1}{2}H_7$ taken off of expansion of the $\beta \cdot H$ expansion. That is,
\begin{equation}
\beta \cdot H=\frac{1}{4}({K^1}_1+\ldots{K^3}_3)-\frac{3}{4}({K^4}_4+\ldots{K^6}_6)+\frac{1}{\sqrt{2}}R_0
\end{equation}
Substituting this into equation (\ref{groupelement}) we find the appropriately arranged element is
\begin{align}
\nonumber g_A=&\exp(-\ln N_2(\frac{1}{4}({K^1}_1+\ldots{K^3}_3)-\frac{3}{4}({K^4}_4+\ldots{K^6}_6)+\frac{1}{\sqrt{2}}\ln N_2R_0))\\
\nonumber &\exp((1-N_2)E_{\beta})\\
=&\exp(-\ln N_2(\frac{1}{4}({K^1}_1+\ldots{K^3}_3)-\frac{3}{4}({K^4}_4+\ldots{K^6}_6)\\
\nonumber &\exp(N_2^{-\frac{1}{4}}(1-N_2)E_{\beta})\exp(-\frac{1}{\sqrt{2}}\ln N_2R_0)
\end{align}
We find a dilaton given by
\begin{equation}
e^A=N_2^{-\frac{1}{\sqrt{2}}}
\end{equation}
And a line element corresponding to a 2-brane identical to (\ref{F2branelineeelement}). The brane is derived from the gauge field given by
\begin{equation}
{A_{123}^T}_{(1)}=N_2^{-\frac{1}{4}}(1-N_2)
\end{equation}
We complete the change to world volume indices using $({e^{\hat{h}})^1}_1=\ldots({e^{\hat{h}})^3}_3=N_0^{-\frac{1}{4}}$ and find
\begin{equation}
{A_{123}}_{(1)}=N_2^{-1}-1
\end{equation}
We conclude that ${R_1}^{456}$ is the generator associated with the dual of ${F_{\mu\nu}}_1$.
\paragraph{A String}
We wish to find the electric brane associated with the ${R^{a_1a_2}}_2$ representation whose highest weight is the generator ${R_2}^{56} (0 0 0 0 1 2 2)$, whose associated root we take to be $\beta$ in equation (\ref{groupelement}) and we note that $(\beta,\beta)=2$. The lowest weight generator is ${R_2}^{12} (1 2 2 2 2 2 2)$ and the corresponding element in the Cartan sub-algebra is
\begin{align}
\nonumber \beta \cdot H&=H_1+2(H_2+\ldots H_5)+H_6+H_7\\
&=\frac{1}{2}({K^1}_1+{K^2}_2)-\frac{1}{2}({K^3}_3+\ldots{K^6}_6)-\sqrt{2}R_0
\end{align}
The corresponding group element (\ref{groupelement}) is
\begin{align}
\nonumber g_A=&\exp(-\frac{1}{2}\ln N_1(\frac{1}{2}({K^1}_1+{K^2}_2)-\frac{1}{2}({K^3}_3+\ldots{K^6}_6)-{\sqrt{2}}\ln N_1R_0))\\
\nonumber &\exp((1-N_1)E_{\beta})\\
\nonumber =&\exp(-\frac{1}{2}\ln N_1(\frac{1}{2}({K^1}_1+{K^2}_2)-\frac{1}{2}({K^3}_3+\ldots{K^6}_6)\\
&\exp(N_1^{-\frac{1}{2}}(1-N_1)E_{\beta})\exp(\frac{1}{\sqrt{2}}\ln N_1R_0)
\end{align}
We find a dilaton given by
\begin{equation}
e^A=N_1^{\frac{1}{\sqrt{2}}}
\end{equation}
And a line element corresponding to a string
\begin{equation}
ds^2=N_1^{-\frac{1}{2}}(-dy_1^2+dx_2^2)+N_1^{\frac{1}{2}}(dx_3^2+\ldots+dx_6^2)
\end{equation}
The brane is derived from the gauge field given by
\begin{equation}
{A_{12}^T}_{(2)}=N_1^{-\frac{1}{2}}(1-N_1)
\end{equation}
We complete the change to world volume indices using $({e^{\hat{h}})^1}_1=({e^{\hat{h}})^2}_2=N_1^{-\frac{1}{4}}$
\begin{equation}
{A_{12}}_{(2)}=N_1^{-1}-1
\end{equation}
This gauge potential gives rise to a 3-form field strength, which we label ${F_{\mu\nu\rho}}_2$.
\paragraph{A Second String}
Starting with with the representation ${R^{a_1a_2}}_0$ whose highest weight is the generator $R_0^{56} (0 0 0 0 1 2 0)$ whose associated root we take for $\beta$ in equation (\ref{groupelement}), noting that $(\beta,\beta)=2$. The lowest weight generator in this representation is $R_0^{12} (1 2 2 2 2 2 0)$ which corresponds to the element in the Cartan sub-algebra given by
\begin{align}
\nonumber \beta \cdot H&=H_1+2(H_2+\ldots H_5)+H_6\\
&=\frac{1}{2}({K^1}_1+{K^2}_2)-\frac{1}{2}({K^3}_3+\ldots{K^6}_6)+{\sqrt{2}}R_0
\end{align}
The corresponding group element (\ref{groupelement}) is
\begin{align}
\nonumber g_A=&\exp(-\frac{1}{2}\ln N_1(\frac{1}{2}({K^1}_1+{K^2}_2)-\frac{1}{2}({K^3}_3+\ldots{K^6}_6)+{\sqrt{2}}\ln N_1R_0))\\
\nonumber &\exp((1-N_1)E_{\beta})\\
\nonumber =&\exp(-\frac{1}{2}\ln N_1(\frac{1}{2}({K^1}_1+{K^2}_2)-\frac{1}{2}({K^3}_3+\ldots{K^6}_6)\\
&\exp(N_1^{-\frac{1}{2}}(1-N_1)E_{\beta})\exp(-\frac{1}{\sqrt{2}}\ln N_1R_0)
\end{align}
We find a dilaton given by
\begin{equation}
e^A=N_1^{-\frac{1}{\sqrt{2}}}
\end{equation}
And a line element corresponding to a string
\begin{equation}
ds^2=N_1^{-\frac{1}{2}}(-dy_1^2+dx_2^2)+N_1^{\frac{1}{2}}(dx_3^2+\ldots+dx_6^2)
\end{equation}
The brane is derived from the gauge field given by
\begin{equation}
{A_{12}^T}_{(0)}=N_1^{-\frac{1}{2}}(1-N_1)
\end{equation}
We complete the change to world volume indices using $({e^{\hat{h}})^1}_1=({e^{\hat{h}})^2}_2=N_1^{-\frac{1}{4}}$
\begin{equation}
{A_{12}}_{(0)}=N_1^{-1}-1
\end{equation}
We conclude that this is the group element associated with the dual of ${F_{\mu\nu\rho}}_2$.
\paragraph{A Gooseberry String} \label{Gooseberrystring}
We find the electric brane associated with the representation ${R^{a_1a_2}}_1$ with highest weight generator $R_1^{56} (0 0 0 0 1 2 1)$, whose associated root we take for $\beta$ in equation (\ref{groupelement}), noting that $(\beta,\beta)=1$. The lowest weight generator in this representation is $R_1^{12} (1 2 2 2 2 2 1)$ and this corresponds to the Cartan sub-algebra element
\begin{align}
\nonumber \beta \cdot H&=H_1+2(H_2+\ldots H_5)+H_6+\frac{1}{2}H_7\\
&=\frac{1}{2}({K^1}_1+{K^2}_2)-\frac{1}{2}({K^3}_3+\ldots{K^6}_6)
\end{align}
Substituting this expression into equation (\ref{groupelement}) gives the corresponding element
\begin{align}
g_A=&\exp(-\ln N_1(\frac{1}{2}({K^1}_1+{K^2}_2)-\frac{1}{2}({K^3}_3+\ldots{K^6}_6)))\exp((1-N_1)E_{\beta})
\end{align}
There is no dilaton and we find a line element corresponding to a string
\begin{equation}
ds^2=N_1^{-1}(-dy_1^2+dx_2^2)+N_1(dx_3^2+\ldots+dx_6^2)
\end{equation}
The brane is derived from the gauge field given by
\begin{equation}
{A_{12}^T}_{(1)}=(1-N_1)
\end{equation}
We complete the change to world volume indices using $({e^{\hat{h}})^1}_1=({e^{\hat{h}})^2}_2=N_1^{-\frac{1}{2}}$
\begin{equation}
{A_{12}}_{(1)}=N_1^{-1}-1
\end{equation}
This gauge field is associated with a second 3-form field strength which we label, ${F_{\mu\nu\rho}}_1$. We deduce that this field strength is self-dual as we have no more two-form generators in (\ref{Fgenerators}) that could produce an electric brane that would be associated with a dual to ${F_{\mu\nu\rho}}_1$. We note that the difference in the line element for the string here and those for the other strings derived from $F_4^{+++}$ is due to the dilaton coupling constant being zero here.
\paragraph{A 3-Brane} \label{3-brane}
The representation $R_1^{a_1a_2a_3a_4}$ has highest weight generator $R_1^{3456}$ $(0012341)$ whose associated root we take as $\beta$ in equation (\ref{groupelement}), we note that $(\beta,\beta)=1$. The lowest weight generator in this representation is $R_1^{1234} (1234441)$. This corresponds to an element in the Cartan sub-algebra given by
\begin{align}
\nonumber \beta \cdot H=&H_1+2H_2+3H_3+4(H_4+H_5)+2H_6+\frac{1}{2}H_7\\
&=0({K^1}_1+\ldots{K^4}_4)-({K^5}_5+{K^6}_6)+{\sqrt{2}}R_0
\end{align}
The group element given by equation (\ref{groupelement}) is
\begin{align}
\nonumber g_A=&\exp(-\ln N_3(0({K^1}_1+\ldots{K^4}_4)-({K^5}_5+{K^6}_6)+{\sqrt{2}}R_0))\exp((1-N_3)E_\beta)\\
=&\exp(-\ln N_3(0({K^1}_1+\ldots{K^4}_4)-({K^5}_5+{K^6}_6)))\exp(N_3^{-1}(1-N_3)E_\beta)\\
\nonumber &\exp(-{\sqrt{2}}\ln N_3R_0)
\end{align}
We find a dilaton given by
\begin{equation}
e^A=N_3^{-{\sqrt{2}}}
\end{equation}
And a line element corresponding to a 3-brane
\begin{equation}
ds^2=(-dy_1^2+\ldots dx_4^2)+N_3^{2}(dx_5^2+dx_6^2)
\end{equation}
The associated gauge potential is
\begin{equation}
{A_{1234}}^T_{(1)}=N_3^{-1}-1
\end{equation}
The expression is the same in world volume indices as $({e^{\hat{h}})^1}_1=\ldots ({e^{\hat{h}})^4}_4=1$ that is
\begin{equation}
{A_{1234}}_{(1)}=N_3^{-1}-1
\end{equation}
This gauge potential gives rise to a 5 form field strength and we conclude this is the dual to the one form field strength formed from $R_1$, ${F_\mu}_1=\partial_\mu R_1$.\\

We have have successfully reproduced all the usual BPS electric branes using the group element given in equation (\ref{groupelement}). The formulae we have found above do indeed correspond to the solutions of the Lagrangian (\ref{Flagrangian}).\\

We find our dilaton field $A$ to be related to $\phi$ by

\begin{equation}
A = -\phi
\end{equation}

\section{Higher Level Branes}
In this paper we generalised the result of \cite{West}, namely that the group element of equation (\ref{groupelement}) encodes the usual electric BPS brane solutions of the $\cal G^{+++}$ theories at low levels. So far we have only considered the low-order field content, up to the level of the dual gravity field. We have found precise agreement with the solutions of the known actions of the oxidised theories \cite{CremmerJuliaLuPope}. The $pp$-wave for each $\cal G^{+++}$ theory is also present in the low order theory and can be found as advocated in \cite{West}. However, as explained in \cite{West} we can also apply equation (\ref{groupelement}) to the higher order generators of any $\cal G^{+++}$ to find putative solutions solutions to the non-linearly realised theory.\footnote{Actually in reference \cite{West} one should have used the formula (\ref{groupelement}) given here to investigate the higher order brane solutions, as the root $\beta$ associated with any higher order generator will generally not satisfy $(\beta,\beta)=2$. Fortunately, for all but the brane associated with the generator $R^{a_1a_2a_3}_{b_1b_2}$, the higher order solutions considered in that paper have $(\beta,\beta)=2$.} In this paper we do not show that these are actually solutions but the elegance of equation (\ref{groupelement}) leads us to hope that they are.\\

Let us consider the example for the $F_4^{+++}$ theory. Field content at higher levels in $F_4^{+++}$ is given in \cite{KleinschmidtSchnakenburgWest}, and we find the field ${A_{a_1a_2a_3a_4}}_3$ at level $(3,4)$. This representation has a highest weight generator $R^{3456}_3$ $(0012343)$ and commutes with the dilaton via $[R_0,R^{a_1a_2a_3a_4}_3]=-\frac{1}{\sqrt{2}}R^{a_1a_2a_3a_4}_3$. Applying our method for the root $\beta$ associated with this generator, for which we note that $(\beta,\beta)=1$, we find a group element from (\ref{groupelement}) given by
\begin{align}
g_A=&\exp(-\ln N_3(0({K^1}_1+\ldots {K^4}_4)-({K^5}_5+{K^6}_6)))\\
\nonumber &\exp(N_3^{-1}(1-N_3)E_\beta)\exp(\sqrt{2}\ln N_3R_0)
\end{align}
A dilaton given by
\begin{equation}
e^A=N_3^{\sqrt{2}}
\end{equation}
A line element corresponding to a 3-brane
\begin{equation}
ds^2=(-dy_1^2+\ldots dx_4^2)+N_3(dx_5^2+dx_6^2)
\end{equation}
The brane is derived from a gauge potential
\begin{equation}
{A_{1234}}_{(3)}^T=N_3^{-1}-1={A_{1234}}_{(3)}
\end{equation}
This gives rise to a 5-form field strength, ${F_{\mu\nu\rho\sigma\tau}}_3$. If the dual field strength were to exist we would expect it to be a 1-form formed from a scalar, which we label $R_{-1}$. Such a scalar does not appear in the field content tables of \cite{KleinschmidtSchnakenburgWest} and does not exist but had it done it would commute with the dilaton via $[R_0, R_{-1}]=\frac{1}{\sqrt{8}}R_{-1}$.\\

As is well known, the eleven dimensional supergravity theory has a $pp$-wave, a 2-brane and a 5-brane whose corresponding conserved charges occur in the eleven dimensional supersymmetry algebra. This is consistent with the results of \cite{AczarragaGauntlettIzquierdoTownsend} which showed that each of the brane solutions had a topological charge that appeared as a central charge in the supersymmetry algebra. In reference \cite{West3} it was argued that the brane solutions of the non-linearly realised $\cal G^{+++}$ theory possessed charges which belonged to the $l_1$ representation of $\cal G^{+++}$. The $l_1$ representation is the fundamental representation of $\cal G^{+++}$ associated with the very-extended node on the corresponding Dynkin diagram. The $l_1$ multiplet of charges for $E_8^{+++}$ begins with the generators of spacetime translations, $P^a$, and then contains the 2-form and the 5-form central charges of the supersymmetry algebra at the lowest levels, as well as an infinite number of other charges in the higher orders. It was shown in \cite{KleinschmidtWest} that the $l_1$ representation has the correct $A_n$ representation to be interpreted as central charges. Indeed the reader may verify this correspondence for the brane solutions found in this paper for each $\cal G^{+++}$ considered. Let us consider the example of $F_4^{+++}$ and list the fields, $A_{[n]}$, with their corresponding conserved charges, $Z^{[n-1]}$, associated with the brane solution we have found in each case, including the higher level field, ${A_{a_1a_2a_3a_4}}_{(s=3)}$, which we have just considered\\
\begin{center}
\begin{tabular}{c|c|c}
Levels, $(l_1,l_2)$ & Fields, $A_{[n]}$ & Associated Charges, $Z^{[n-1]}$\\
\hline
$(0,0)$ & ${\hat{h}}_a\hspace{0pt}^b$ & $P^a$\\
$(0,0)$ & $A$ & $-$\\
$(1,0)$ & $A_1$ & $-$\\
$(0,1), (1,1)$ & $A_{a_1}^i$ & $Z^i$\\
$(0,2),(1,2),(2,2)$ & $A_{a_1a_2}^{(ij)}$ & $Z^{a_1(ij)}$\\
$(1,3),(2,3)$ & $A_{a_1a_2a_3}^i$ & $Z^{a_1a_2i}$\\
$(2,4)$ & $A_{a_1a_2a_3,b}$ & $Z^{a_1a_2,b}$\\
$(1,4),(2,4),(3,4)$ & $A_{a_1a_2a_3a_4}^{(ij)}$ & $Z^{a_1a_2a_3(ij)}$\\
\end{tabular}
\end{center}
\vspace{5pt}
The $F_4^{+++}$ theory possesses an $SL(2)$ symmetry associated with the seventh node of its Dynkin diagram, which has the generator $R_1$, which gives rise to the assignment of the $i,j$ indices above.
We note that while we could not use our method to find the brane solution associated to ${A_{a_1a_2a_3a_4}}_{(s=2)}$, as its generator had associated root, $\beta$, such that $(\beta,\beta)=0$, we are able to deduce that its field strength is dual to a 1-form field strength formed from the dilaton, $A$. Indeed we expect its brane solution to have a third rank tensor conserved charge, as listed above.\\

It is interesting to compare these with the supersymmetric central charges of the $(1,0)$ supersymmetry algebra
\begin{equation}
\left\{Q_\alpha^i,Q_\beta^j\right\}=\epsilon^{ij}(\Gamma_m)_{\alpha\beta}P^m+(\Gamma_{m_1m_2m_3})_{\alpha\beta}Z^{m_1m_2m_3(ij)} \label{SuSy}
\end{equation}
The supersymmetric generator of spacetime translations, $P^m$, is equivalent to the conserved charge arising from the $pp$-wave solution corresponding to the graviton, ${\hat{h}}_a\hspace{0pt}^b$, in the $F_4^{+++}$ field content. Similarly we can make an association between the central charge $Z^{m_1m_2m_3(ij)}$ of the supersymmetry algebra and the conserved charges coming from brane solutions which couple to the fields $A_{a_1a_2a_3a_4}^i$. The 5-form field strengths constructed from ${A_{a_1a_2a_3a_4}}_{(s=1)}$ and ${A_{a_1a_2a_3a_4}}_{(s=2)}$ are dual to the 1-form field strengths formed from the axion, $A_1$, and the dilaton, $A$, respectively, which can be seen by referring to the 3-brane solution found on page \pageref{3-brane} and from considering the low-level content. We have just considered, above in this section, the field, ${A_{a_1a_2a_3a_4}}_{(s=3)}$, whose field strength is a 5-form and its associated brane solution is also a 3-brane. Altogether we have a triplet of conserved charges coming from the non-linear realisation of $F_4^{+++}$ which are equivalent to the central charges, $Z^{m_1m_2m_3(ij)}$, of the supersymmetry algebra. Including the generators of spacetime translations, $P^m$, the conserved charges, $Z^{m_1m_2m_3(ij)}$, of $F_4^{+++}$ account for 36 degrees of freedom which is compatible with the anticommutator of the two $Sp(2)$ Majorana-Weyl spinors, $Q_\alpha^i$.\\
 
In this paper we have constructed branes, even at a lower level than those mentioned above, whose conserved charges in the above table are not present amongst the supersymmetry algebra's central charges; charges such as $Z^i$ and the part of $Z^{a_1(ij)}$ triplet associated with the so-called "gooseberry string", listed in our analysis of $F_4^{+++}$ on page \pageref{Gooseberrystring}. Indeed we can carry on and consider the brane solutions associated with the infinite number of higher-order fields in $F_4^{+++}$, or any $\cal G^{+++}$, and find a conserved charge for each one. We therefore find that the central charges in the supersymmetry algebra only account for a few of the brane charges, all of which belong to the $l_1$ representation.

{\centering \section*{Acknowledgements}}
We thank Andrew Pressley for useful conversations. This research was supported by the PPARC grants PPA/S/S/2003/03644 and PPA/Y/S/2002/001/44.\\

This paper extends the considerations of \cite{West} to find a universal formula for the $\cal G^{+++}$ elements which expresses the usual BPS solutions for single branes found in $\cal G^{+++}$ at low orders. While this paper was in the final stages of preparation a paper with related results \cite{EnglertHouart1} was submitted to the archive.

\newpage
\appendix
\section{The Electric Branes of $E_6^{+++}$, $E_7^{+++}$ and $G_2^{+++}$}
\subsection{Very Extended $E_6$}
The Dynkin diagram for $E_6^{+++}$ is given in Appendix B. The $E_6^{+++}$ algebra may only be uniquely decomposed with respect to an $A_7$ gravity line, giving an 8-dimensional theory. The simple roots that we eliminate are $\alpha_9$ and $\alpha_8$ whose corresponding generators are $R_1 (00\ldots 001)$ and $R_0^{678} (00\ldots 010)$ and we associate the levels $l_1$ and $l_2$ with these respectively. The $s$ labels on our generators are chosen to be the $l_1$ level of the generator. The alternative decomposition along the other branch of the Dynkin diagram is related to the first decomposition via an automorphism.\\

From reference \cite{KleinschmidtSchnakenburgWest} we find the remaining algebra contains the generators ${K^a}_b$ at level (0,0) and the following generators at up to level (1,2)
\begin{alignat}{3}
\nonumber l_1 \longrightarrow \qquad &\qquad &\qquad &0 &\qquad &1\\ 
\nonumber l_2 \downarrow \qquad \qquad &0 &\qquad & R_0 & \qquad & R_1\\ 
&1 &\qquad &R_0^{a_1a_2a_3} & \qquad &R_1^{a_1a_2a_3} \label{E6generators}\\
\nonumber &2 &\qquad &R_0^{a_1\ldots a_6} & \qquad &R_1^{a_1\ldots a_6}\\
\nonumber & &\qquad & & \qquad &R_1^{a_1\ldots a_5,b}
\end{alignat}
All the generators have $(\beta,\beta)=2$, where $\beta$ is the associated root for the generator, with the exceptions of $R_0$ and $R_1^{a_1\ldots a_6}$ which have $(\beta,\beta)=0$ and so will not be used as starting points for finding electric branes.\\

We make use of the following commutator relations where we have chosen the commutator coefficient for $[R_0,R_1]$ and the others have followed from equations (\ref{jicoefficients1})-(\ref{jicoefficients4}) and the Serre relations (\ref{Serre})
\begin{alignat}{3}
\nonumber \qquad \qquad & &\qquad & & \qquad & [R_0,R_1]=-R_1\\ 
& &\qquad &[R_0,R_0^{a_1a_2a_3}]=\frac{1}{2}R_0^{a_1a_2a_3} & \qquad &[R_0,R_1^{a_1a_2a_3}]=-\frac{1}{2}{R_1}^{a_1a_2a_3}\\
\nonumber & &\qquad &[R_0,R_0^{a_1\ldots a_6}]=R_0^{a_1\ldots a_6} & \qquad &[R_0,R_1^{a_1\ldots a_6}]=0
\end{alignat}
The simple root generators of $E_6^{+++}$ are
\begin{equation}
E_a={K^a}_{a+1}, a=1,\ldots 7, E_8=R_0^{678}, E_9=R_1
\end{equation}
and the Cartan sub-algebra generators, $H_a$, are given by
\begin{eqnarray}
\nonumber &H_a={K^a}_a-{K^{a+1}}_{a+1}, a=1,\ldots 7,\\
&H_8=-\frac{1}{2}({K^1}_1+\ldots{K^5}_5)+\frac{1}{2}({K^6}_6+\ldots{K^8}_8)+R_0,\\
\nonumber&H_9=-2R_0
\end{eqnarray}
The low-level field content \cite{KleinschmidtSchnakenburgWest} is ${\hat{h}}_a\hspace{0pt}^b$, $A$, ${A_{a_1a_2a_3}}_0$, ${A_{a_1\ldots a_6}}_0$ and their field strengths have duals derived from  ${A_{a_1\ldots a_5,b}}_1$, ${A_{a_1\ldots a_6}}_1$, ${A_{a_1a_2a_3}}_1$, $A_1$, respectively. Our choice of local sub-algebra for the non-linear realisation allows us to write the group element of $E_6^{+++}$ as
\begin{align}
\nonumber g=\exp(\sum_{a\leq b}{\hat{h}}_a\hspace{0pt}^b{K^a}_b)&\exp(\frac{1}{6!}{A_{a_1\ldots a_6}}_1R_1^{a_1\ldots a_6})\exp(\frac{1}{6!}{A_{a_1\ldots a_6}}_0R_0^{a_1\ldots a_6})\\
&\exp(\frac{1}{5!}{A_{a_1\ldots a_5,b}}_1R_1^{a_1\ldots a_5,b})\exp(\frac{1}{3!}{A_{a_1a_2a_3}}_1R_1^{a_1\ldots a_3})\\
\nonumber &\exp(\frac{1}{3!}{A_{a_1a_2a_3}}_0R_0^{a_1\ldots a_3})\exp(A_1 R_1)\exp(AR_0)
\end{align}
The field content of the associated maximally oxidised theory given in \cite{CremmerJuliaLuPope} agrees with the low-order $E_6^{+++}$ content given above. The Lagrangian for the oxidised theory, contains contains a graviton, $\phi$, a 4-form field strength, ${F_{\mu\nu\rho\sigma}}_0=4\partial_{[\mu}{A_{\nu\rho\sigma]}}_0$, a 1-form field strength, ${F_{\mu}}_1=\partial_{\mu}A_1$ and a dilaton $A$ and is given by
\begin{align}
\nonumber A=\frac{1}{16\pi G_{n-1}}\int d^8x\sqrt{-g}(&R-\frac{1}{2}\partial_\mu\phi\partial^\mu\phi-\frac{1}{2}e^{2\phi}{F_{\mu}}_1F^{\mu}_1\\
&-\frac{1}{2.4!}e^{-\phi}{F_{\mu\nu\rho\sigma}}_0F^{\mu\nu\rho\sigma}_0)\\
\nonumber +\int_{C.S.} d^8x\epsilon^{a_1\ldots a_8}&\frac{1}{4!4!}A_1{F_{a_1\ldots a_4}}_0{F_{a_5\ldots a_8}}_0
\label{E6lagrangian}
\end{align}
Using the group element (\ref{groupelement}) we find the following electric branes
\paragraph{A 2-Brane}
Let us find the electric brane associated with the representation ${R^{a_1a_2a_3}}_0$ whose highest weight generator is $R_0^{678}$ $(00\ldots010)$, whose associated root we take as $\beta$ in equation (\ref{groupelement}). The lowest weight generator in this representation is $R_0^{123}$ $(123\ldots32110)$ and the associated Cartan sub-algebra element is
\begin{align}
\nonumber \beta \cdot H=&H_1+2H_2+3(H_3+\ldots H_5)+2H_6+H_7+H_8\\
=&\frac{1}{2}({K^1}_1+\ldots{K^3}_3)-\frac{1}{2}({K^4}_4+\ldots{K^8}_8)+R_0
\end{align}
Bearing in mind that $(\beta,\beta)=2$, we find that the group element (\ref{groupelement}) is
\begin{align}
\nonumber g_A=&\exp(-\frac{1}{2}\ln N_2(\frac{1}{2}({K^1}_1+\ldots{K^3}_3)-\frac{1}{2}({K^4}_4+\ldots{K^8}_8)))\\
&\exp(N_2^{-\frac{1}{4}}(1-N_2)E_{\beta})\exp(-\frac{1}{2}\ln N_2R_0)
\end{align}
We have a dilaton given by
\begin{equation}
e^A=N_2^{-\frac{1}{2}}
\end{equation}
And a line element corresponding to a 2-brane
\begin{equation}
ds^2=N_2^{-\frac{1}{2}}(-dy_1^2+dx_2^2+dx_3^2)+N_2^{\frac{1}{2}}(dx_4^2+\ldots+dx_8^2)
\end{equation}
The brane is derived from a gauge potential
\begin{equation}
{A_{123}^T}_{(0)}=N_2^{-\frac{1}{4}}(1-N_2)
\end{equation}
The change to world volume indices is made using $({e^{\hat{h}})^1}_1=\ldots({e^{\hat{h}})^3}_3=N_2^{-\frac{1}{4}}$
\begin{equation}
{A_{123}}_{(0)}=N_2^{-1}-1
\end{equation}
This gauge potential gives rise to a 4-form field strength, ${F_{\mu\nu\rho\sigma}}_0$.
\paragraph{A Second 2-Brane}
Let us find the brane associated with the representation ${R^{a_1}}_1$ whose highest weight generator is $R_1^{678}$ $(00\ldots011)$, whose associated root we take for $\beta$ in equation (\ref{groupelement}). The lowest weight generator in this representation is $R_1^{123}$ $(123\ldots32111)$ and the corresponding element in the Cartan sub-algebra is identical to that for $R_0^{123}$ but with $2R_0$ taken off, that is,
\begin{equation}
\beta \cdot H=\frac{1}{2}({K^1}_1+\ldots{K^3}_3)-\frac{1}{2}({K^4}_4+\ldots{K^8}_8)-R_0
\end{equation}
We find that the group element of equation (\ref{groupelement}) is
\begin{align}
\nonumber g_A=&\exp(-\frac{1}{2}\ln N_2(\frac{1}{2}({K^1}_1+\ldots{K^3}_3)-\frac{1}{2}({K^4}_4+\ldots{K^8}_8)))\\
&\exp(N_2^{-\frac{1}{4}}(1-N_2)E_{\beta})\exp(\frac{1}{2}\ln N_2R_0)
\end{align}
We have a dilaton given by
\begin{equation}
e^A=N_2^{\frac{1}{2}}
\end{equation}
And a line element corresponding to a 2-brane
\begin{equation}
ds^2=N_2^{-\frac{1}{2}}(-dy_1^2+dx_2^2+dx_3^2)+N_2^{\frac{1}{2}}(dx_4^2+\ldots+dx_8^2)
\end{equation}
The brane is derived from a gauge potential
\begin{equation}
{A_{123}^T}_{(1)}=N_2^{-\frac{1}{4}}(1-N_2)
\end{equation}
The change to world volume indices is made using $({e^{\hat{h}})^1}_1=\ldots({e^{\hat{h}})^3}_3=N_2^{-\frac{1}{4}}$
\begin{equation}
{A_{123}}_{(1)}=N_2^{-1}-1
\end{equation}
This gives rise to a 4-form field strength, which we conclude is the dual to ${F_{\mu\nu\rho\sigma}}_0$.
\paragraph{A 5-Brane}
Let us find the electric brane associated with the representation ${R^{a_1a_2\ldots a_6}}_0$ which has highest weight generator $R_0^{345678}(001232120)$, whose associated root we take as $\beta$ in equation (\ref{groupelement}). The lowest weight generator in this representation is $R_0^{123456}(123454220)$ and the corresponding element in the Cartan sub-algebra is
\begin{align}
\nonumber \beta \cdot H=&H_1+2H_2+3H_3+4H_4+5H_5+4H_6+2H_7+2H_8\\
=&0({K^1}_1+\ldots{K^6}_6)-({K^7}_7+{K^8}_8)+2R_0
\end{align}
The group element (\ref{groupelement}) is
\begin{align}
\nonumber g_A=&\exp(\frac{1}{2}\ln N_5({K^7}_7+{K^8}_8))\\
&\exp(N_5^{-1}(1-N_5)E_{\beta})\exp(-\ln N_5R_0)
\end{align}
We have a dilaton given by
\begin{equation}
e^A=N_5^{-1}
\end{equation}
And a line element corresponding to a 5-brane
\begin{equation}
ds^2=(-dy_1^2+dx_2^2+\ldots dx_6^2)+N_5(dx_7^2+dx_8^2)
\end{equation}
The brane is derived from a gauge potential
\begin{equation}
{A_{12\ldots 6}^T}_{(0)}=N_5^{-1}-1
\end{equation}
The change to world volume indices leaves the form of the gauge potential unaltered as $({e^{\hat{h}})^1}_1=\ldots({e^{\hat{h}})^6}_6=1$
\begin{equation}
{A_{12\ldots 6}}_{(0)}=N_5^{-1}-1
\end{equation}
This gauge potential gives rise to a 7-form field strength, which we conclude is the dual to the 1 form ${F_\mu}_1=\partial_\mu R_1$.\\

We have have successfully reproduced all the usual BPS electric branes using the group element given in equation (\ref{groupelement}). The formulae we have found above do indeed correspond to the solutions of the Lagrangian (\ref{E6lagrangian}).\\

We find our dilaton field $A$ to be related to $\phi$ by

\begin{equation}
A = \phi
\end{equation}

\subsection{Very Extended $E_7$ - The 10-dimensional theory}
The $E_7^{+++}$ algebra may be decomposed with respect to an $A_9$ gravity line, giving a 10-dimensional theory or equally with respect to its $A_7$ gravity line to give an 8-dimensional theory. We consider the 10-dimensional theory here, and note that in doing so we depart from the considerations of \cite{CremmerJuliaLuPope}, in which the 9-dimensional theory is considered.\\

The simple root whose node we delete is $\alpha_{10}$ and has generator $R^{78910}$ $(00\ldots 001)$ and we associate the level $l$ with it. Our $s$ labels are all zero, as we eliminate only one simple root, and are discarded in this section. From reference \cite{KleinschmidtSchnakenburgWest} we find the ${E_7}^{+++}$ algebra decomposed with respect to an $A_9$ sub-algebra contains the generators ${K^a}_b$ at level 0 and the following generators up to level 2, and notably no dilaton generator,
\begin{alignat}{3}
\nonumber l \downarrow \qquad & &\qquad & &\qquad & \\
\nonumber 1 \qquad & &\qquad &R^{a_1a_2a_3a_4} &\qquad & \\
 2 \qquad & &\qquad &R^{a_1\ldots a_7,b} &\qquad & \label{E7generators}\\
\nonumber \qquad & &\qquad &R^{a_1\ldots a_8} &\qquad &
\end{alignat}
We note that the generator $R^{a_1\ldots a_8}$ has associated root $\beta$ such that $(\beta,\beta)=0$ so we discard this as a starting point for finding an electric brane, all the other generators have $(\beta,\beta)=2$.
The simple root generators of $E_7^{+++}$ are
\begin{equation}
E_a={K^a}_{a+1}, a=1,\ldots 9, E_{10}=R^{78910}
\end{equation}
and the Cartan sub-algebra generators, $H_a$, are given by
\begin{eqnarray}
\nonumber &H_a={K^a}_a-{K^{a+1}}_{a+1}, a=1,\ldots 9,\\
&H_{10}=-\frac{1}{2}({K^1}_1+\ldots{K^6}_6)+\frac{1}{2}({K^7}_7+\ldots{K^{10}}_{10})
\end{eqnarray}
The low-level field content \cite{KleinschmidtSchnakenburgWest} is ${\hat{h}}_a\hspace{0pt}^b$, and its field strength has a dual derived from  ${A_{a_1\ldots a_7,b}}$, we also have fields ${A_{a_1a_2a_3a_4}}$, ${A_{a_1\ldots a_8}}$ which are not related to each other by a duality condition and we take them to be self-dual in our low order theory. Our choice of local sub-algebra for the non-linear realisation allows us to write the group element of $E_{7}^{+++}$ as
\begin{align}
\nonumber g=\exp(\sum_{a\leq b}{\hat{h}}_a\hspace{0pt}^b{K^a}_b)&\exp(\frac{1}{8!}{A_{a_1\ldots a_8}}R^{a_1\ldots a_8})\exp(\frac{1}{7!}A_{a_1\ldots a_7,b}R^{a_1\ldots a_7,b})\\
&\exp(\frac{1}{4!}A_{a_1\ldots a_4}R^{a_1\ldots a_4})
\end{align}
\subsubsection{A 3-Brane}
Let us find the electric brane associated with the representation $R^{a_1a_2a_3a_4}$ whose highest weight generator is $R^{78910}$ $(0 0\ldots 0 0 1)$, whose root, $\beta$, we use in equation (\ref{groupelement}). The lowest weight generator in this representation is $R^{1234}$ $(1 2 3 4 4 4 3 2 1 1)$ which corresponds to the Cartan sub-algebra element given by
\begin{align}
\nonumber \beta \cdot H=&H_1+2H_2+3H_3+4(H_4+\ldots H_6)+3H_7+2H_8+H_9+H_{10}\\
=&\frac{1}{2}({K^1}_1+\ldots{K^4}_4)-\frac{1}{2}({K^5}_5+\ldots{K^{10}}_{10})
\end{align}
We now write down the group element from equation (\ref{groupelement})
\begin{equation}
g_A=\exp(-\frac{1}{2}\ln N_3(\frac{1}{2}({K^1}_1+\ldots{K^4}_4)-\frac{1}{2}({K^5}_5+\ldots{K^{10}}_{10})))\exp((1-N_3)E_{\beta})
\end{equation}
We have a line element corresponding to a 3-brane
\begin{equation}
ds^2=N_3^{-\frac{1}{2}}(-dy_1^2+dx_2^2+\ldots dx_4^2)+N_3^{\frac{1}{2}}(dx_5^2+\ldots+dx_{10}^2)
\end{equation}
The brane is derived from a gauge potential
\begin{equation}
A_{1234}^T=1-N_3
\end{equation}
We complete the change to world volume indices using $({e^{\hat{h}})^1}_1=\ldots({e^{\hat{h}})^4}_4=N_3^{-\frac{1}{4}}$
\begin{equation}
A_{1234}=N_3^{-1}-1
\end{equation}
This gives rise to a 5-form field strength,$F_{\mu\nu\rho\sigma\tau}$ which we conclude is self-dual, as there are no other other generators in (\ref{E7generators}) from which we could derive a dual field strength, and consequently we cannot construct a Lagrangian for this theory.
\subsection{Very Extended $G_2$}
The Dynkin diagram for ${G_2}^{+++}$ is shown in Appendix B, with the darkened nodes indicating the gravity line we consider here. The ${G_2}^{+++}$ algebra is decomposed with respect to the $A_4$ sub-algebra of the gravity line where the deleted node corresponds to simple root is $\alpha_5$ whose generator is $R^5 (00001)$, to which our decomposition level, $l$, is associated. Again we have no need for the $s$ labels as they are all zero, so we discard them in this section. Reference \cite{KleinschmidtSchnakenburgWest} gives the generators at low levels, we find the ${K^a}_b$ at the zeroth level corresponding to the gravity line step operators and we have the following other generators up to level 3, and notably no dilaton generator,
\begin{alignat}{3}
\nonumber l \downarrow \qquad & &\qquad & &\qquad & \\
\nonumber 1 \qquad & &\qquad &R^a &\qquad & \\
 2 \qquad & &\qquad &R^{ab} &\qquad & \label{G2generators}\\
\nonumber 3 \qquad & &\qquad &R^{ab,c} &\qquad &\\
\nonumber  \qquad & &\qquad &R^{abc} &\qquad &
\end{alignat}
The generators $R^a$, $R^{ab}$ and $R^{ab,c}$ each have an associated root $\beta$ such that $(\beta,\beta)=\frac{2}{3}$ but $R^{abc}$ has $(\beta,\beta)=0$ so we shall discard it as a starting point for our method of finding electric branes.\\

The simple root generators of ${G_2}^{+++}$ are
\begin{equation}
E_a={K^a}_{a+1}, a=1,\ldots4, E_5=R^5
\end{equation}
and the Cartan sub-algebra generators, $H_a$, are given by
\begin{eqnarray}
\nonumber &H_a={K^a}_a-{K^{a+1}}_{a+1}, a=1,\ldots4,\\
&H_5=-({K^1}_1+\ldots{K^4}_4)+2{K^5}_5
\end{eqnarray}
The low-level field content \cite{KleinschmidtSchnakenburgWest} is ${\hat{h}}_a\hspace{0pt}^b$, $A_{a_1}$ and their field strengths have duals derived from  $A_{a_1a_2,b}$, $A_{a_1a_2}$, respectively. We also find the field $A_{a_1a_2a_3}$ which is not related to any of the other fields by a duality condition and we take it to be self-dual in our low order theory. Our choice of local sub-algebra for the non-linear realisation allows us to write the group element of $G_2^{+++}$ as
\begin{align}
g=\exp(\sum_{a\leq b}{\hat{h}}_a\hspace{0pt}^b{K^a}_b)&\exp(\frac{1}{3!}A_{a_1a_2a_3}R^{a_1a_2a_3})\exp(\frac{1}{2!}A_{a_1a_2,b}R^{a_1a_2,b})\\
\nonumber &\exp(\frac{1}{2!}A_{a_1a_2}{R}^{a_1a_2})\exp(A_{a_1}R^{a_1})
\end{align}
The field content of the associated maximally oxidised theory given in \cite{CremmerJuliaLuPope} agrees with the low-order $G_2^{+++}$ content given above. The Lagrangian for the oxidised theory, contains contains a graviton, $\phi$, and contains a 2-form field strength $F_{\mu\nu}=2\partial_{[\mu}A_{\nu]}$, and is given by
\begin{equation}
A=\frac{1}{16\pi G_5}\int d^5x\sqrt{-g}(R-\frac{1}{2.2!}F_{\mu\nu}F^{\mu\nu})+\int_{C.S.} d^5x\epsilon^{a_1\ldots a_5}\frac{1}{3!2!\sqrt{3}}F_{a_1a_2}F_{a_3a_4}A_{a_5} \label{G2lagrangian}
\end{equation}
We find the following electric branes of $G_2^{+++}$ via our group element (\ref{groupelement})
\paragraph{A Particle}
We find the highest weight generator associated with the representation $R^{a_1}$ is $R^5 (00001)$, whose associated root we take for $\beta$ in equation (\ref{groupelement}), noting that $(\beta,\beta)=\frac{2}{3}$. We find that the lowest weight generator in this representation is $R^1 (11111)$. Taking account that the fifth simple root, as enumerated in Appendix B, of ${G_2}^{+++}$ is shorter than the other four such that $\frac{(\alpha_a,\alpha_a)}{(\alpha_5,\alpha_5)}=3$ where $a=1\ldots 4$ we find that the expansion of $R^1$ in the Cartan sub-algebra according to (\ref{Cartanbasis}) is
\begin{align}
\nonumber \beta \cdot H=&H_1+H_2+H_3+H_4+\frac{1}{3}H_5\\
&=\frac{2}{3}({K^1}_1)-\frac{1}{3}({K^2}_2+\ldots{K^5}_5)
\end{align}
The group element in equation (\ref{groupelement}) takes the form
\begin{equation}
g_A=\exp(-\frac{3}{2}\ln N_0(\frac{2}{3}({K^1}_1)-\frac{1}{3}({K^2}_2+\ldots{K^5}_5)))\exp((1-N_0)E_{\beta})
\end{equation}
We read off the line element of a particle
\begin{equation}
ds^2=N_0^{-2}(-dy_1^2)+N_0(dx_2^2+\ldots+dx_5^2)
\end{equation}
The associated gauge potential is read from the group element to be
\begin{equation}
{A_1}^T=(1-N_0)
\end{equation}
We complete the change to world volume indices using $({e^{\hat{h}})^1}_1=N_0^{-1}$
\begin{equation}
A_1=N_0^{-1}-1
\end{equation}
This gauge potential gives rise to a 2-form field strength, $F_{\mu\nu}$.
\paragraph{A String}
The highest weight generator in the representation $R^{a_1a_2}$ is $R^{45}$ $(00012)$ whose associated root $\beta$ we take for $\beta$ in equation (\ref{groupelement}). The lowest weight generator in this representation is $R^{12} (12222)$ and, noting that $(\beta,\beta)=\frac{2}{3}$, we find that the corresponding element in the Cartan sub-algebra is
\begin{align}
\nonumber \beta \cdot H=&H_1+2(H_2+\ldots H_4)+\frac{2}{3}H_5\\
&=\frac{1}{3}({K^1}_1+{K^2}_2)-\frac{2}{3}({K^3}_3+\ldots{K^5}_5)
\end{align}
The group element of equation (\ref{groupelement}) takes the form
\begin{equation}
g_A=\exp(-\frac{3}{2}\ln N_1(\frac{1}{3}({K^1}_1+{K^2}_2)-\frac{2}{3}({K^3}_3+\ldots{K^5}_5)))\exp((1-N_1)E_{\beta})
\end{equation}
We have the line element of a string
\begin{equation}
ds^2=N_1^{-1}(-dy_1^2+dx_2^2)+N_1^{2}(dx_3^2+\ldots+dx_5^2)
\end{equation}
The form of the line element is the same as that of the brane associated with the dual of $F_{\mu\nu}$, the 2-form field strength which we obtained in the previous section. The associated gauge potential is read from the group element to be
\begin{equation}
{A_{12}}^T=(1-N_1)
\end{equation}
We change to world volume indices using $({e^{\hat{h}})^1}_1=({e^{\hat{h}})^2}_2=N_1^{-\frac{1}{2}}$
\begin{equation}
A_{12}=N_1^{-1}-1
\end{equation}
\\
We have have successfully reproduced all the usual BPS electric branes of the oxidised $G_2$ theory using the group element given in equation (\ref{groupelement}). The formulae we have found above do indeed correspond to the solutions of the Lagrangian (\ref{G2lagrangian}).\\
\newpage
\section{Dynkin Diagrams}
\newcounter{cms}
\setlength{\unitlength}{0.8mm}
\subsection{$B_{n-3}^{+++}$}
\begin{picture}(50,40)
\put(5,20){\line(1,0){60}}
\put(5,20){\circle*{10}}
\put(4,12){$1$}
\put(20,20){\circle*{10}}
\put(19,12){$2$}
\put(35,20){\circle*{10}}
\put(34,12){$3$}
\put(50,20){\circle*{10}}
\put(49,12){$4$}
\put(50,20){\line(0,1){8}}
\put(50,31.6){\circle{7}}
\put(43,30){$n$}
\put(55,30){$R_1^{56\ldots (n-1)}$}
\put(65,20){\circle*{10}}
\put(64,12){$5$}
\put(65,20){\line(1,0){5}}
\put(68,13){\ldots}
\put(75,20){\line(1,0){5}}
\put(80,20){\circle*{10}}
\put(75,12){$(n-3)$}
\put(95,20){\circle*{10}}
\put(90,12){$(n-2)$}
\put(93,20){\line(0,1){6}}
\put(97,20){\line(0,1){6}}
\put(90,23){\line(1,1){5}}
\put(100,23){\line(-1,1){5}}
\put(80,20){\line(1,0){15}}
\put(95,31.6){\circle{7}}
\put(78,30){$(n-1)$}
\put(100,30){$R_0^{(n-1)}$}
\end{picture}
\subsection{$D_{n-3}^{+++}$}
\begin{picture}(50,40)
\put(5,20){\line(1,0){60}}
\put(5,20){\circle*{10}}
\put(4,12){$1$}
\put(20,20){\circle*{10}}
\put(19,12){$2$}
\put(35,20){\circle*{10}}
\put(34,12){$3$}
\put(50,20){\circle*{10}}
\put(49,12){$4$}
\put(50,20){\line(0,1){8}}
\put(50,31.6){\circle{7}}
\put(43,30){$n$}
\put(55,30){$R_1^{56\ldots (n-1)}$}
\put(65,20){\circle*{10}}
\put(64,12){$5$}
\put(65,20){\line(1,0){5}}
\put(68,13){\ldots}
\put(75,20){\line(1,0){5}}
\put(80,20){\line(1,0){30}}
\put(80,20){\circle*{10}}
\put(75,12){$(n-4)$}
\put(95,20){\circle*{10}}
\put(90,12){$(n-3)$}
\put(110,20){\circle*{10}}
\put(105,12){$(n-2)$}
\put(95,20){\line(0,1){8}}
\put(95,31.6){\circle{7}}
\put(78,30){$(n-1)$}
\put(100,30){$R_0^{(n-2)(n-1)}$}
\end{picture}
\subsection{$E_6^{+++}$}
\begin{picture}(50,50)
\put(5,20){\line(1,0){90}}
\put(5,20){\circle*{10}}
\put(4,12){$1$}
\put(20,20){\circle*{10}}
\put(19,12){$2$}
\put(35,20){\circle*{10}}
\put(34,12){$3$}
\put(50,20){\circle*{10}}
\put(49,12){$4$}
\put(65,20){\circle*{10}}
\put(64,12){$5$}
\put(80,20){\circle*{10}}
\put(79,12){$6$}
\put(95,20){\circle*{10}}
\put(94,12){$7$}
\put(65,20){\line(0,1){8}}
\put(65,31.6){\circle{7}}
\put(58,30){$8$}
\put(70,30){$R_0^{678}$}
\put(65,35.1){\line(0,1){4.5}}
\put(65,43.2){\circle{7}}
\put(58,41.6){$9$}
\put(70,41.6){$R_1$}
\end{picture}
\subsection{$E_7^{+++}$}
\begin{picture}(50,40)
\put(5,20){\line(1,0){120}}
\put(5,20){\circle*{10}}
\put(4,12){$1$}
\put(20,20){\circle*{10}}
\put(19,12){$2$}
\put(35,20){\circle*{10}}
\put(34,12){$3$}
\put(50,20){\circle*{10}}
\put(49,12){$4$}
\put(65,20){\circle*{10}}
\put(64,12){$5$}
\put(80,20){\circle*{10}}
\put(79,12){$6$}
\put(95,20){\circle*{10}}
\put(94,12){$7$}
\put(110,20){\circle*{10}}
\put(109,12){$8$}
\put(125,20){\circle*{10}}
\put(124,12){$9$}
\put(80,20){\line(0,1){8}}
\put(80,31.6){\circle{7}}
\put(71,30){$10$}
\put(85,30){$R^{78910}$}
\end{picture}
\subsection{$F_4^{+++}$}
\begin{picture}(50,50)
\put(5,20){\line(1,0){60}}
\put(5,20){\circle*{10}}
\put(4,12){$1$}
\put(20,20){\circle*{10}}
\put(19,12){$2$}
\put(35,20){\circle*{10}}
\put(34,12){$3$}
\put(50,20){\circle*{10}}
\put(49,12){$4$}
\put(65,20){\circle*{10}}
\put(64,12){$5$}
\put(63,20){\line(0,1){6}}
\put(67,20){\line(0,1){6}}
\put(60,23){\line(1,1){5}}
\put(70,23){\line(-1,1){5}}
\put(65,31.6){\circle{7}}
\put(58,30){$6$}
\put(70,30){$R_0^6$}
\put(65,35.1){\line(0,1){4.5}}
\put(65,43.2){\circle{7}}
\put(58,43.2){$7$}
\put(70,43.2){$R_1$}
\end{picture}
\subsection{$G_2^{+++}$}
\begin{picture}(50,40)
\put(5,20){\line(1,0){45}}
\put(5,20){\circle*{10}}
\put(4,12){$1$}
\put(20,20){\circle*{10}}
\put(19,12){$2$}
\put(35,20){\circle*{10}}
\put(34,12){$3$}
\put(50,20){\circle*{10}}
\put(49,12){$4$}
\put(48,20){\line(0,1){6}}
\put(52,20){\line(0,1){6}}
\put(45,23){\line(1,1){5}}
\put(55,23){\line(-1,1){5}}
\put(50,31.6){\circle{7}}
\put(43,30){$5$}
\put(55,30){$R^5$}
\end{picture}

\end{document}